\documentclass[fleqn,usenatbib]{mnras}
\usepackage{newtxtext,newtxmath}
\usepackage[T1]{fontenc}
\DeclareRobustCommand{\VAN}[3]{#2}
\let\VANthebibliography\thebibliography
\def\thebibliography{\DeclareRobustCommand{\VAN}[3]{##3}\VANthebibliography}

\usepackage{graphicx}
\usepackage{amsmath}
\usepackage{mathrsfs} 
\usepackage{physics}
\usepackage{comment}
\usepackage{url}
\usepackage{bm}
\newcommand{\ee}{\mathrm{e}}

\title[Resolving CGWs in the Presence of GWB]{Resolving Individual Signals in the Presence of Stochastic Background\\ in Future Pulsar Timing Arrays}

\author[K. Furusawa et al.]{
Kazuya Furusawa,$^{1}$\thanks{E-mail: furusawa.kazuya.m8@s.mail.nagoya-u.ac.jp}
Sachiko Kuroyanagi,$^{2,1}$\thanks{E-mail: sachiko.kuroyanagi@csic.es}
Kiyotomo Ichiki$^{1,3}$\thanks{E-mail: ichiki.kiyotomo.a9@f.mail.nagoya-u.ac.jp}
\\
$^{1}$Graduate School of Science, Nagoya University, Furo-cho, Chikusa-ku, Nagoya, Aichi, 464-8602, Japan\\
$^{2}$Instituto de F\'{i}sica Te\'{o}rica UAM-CSIC, Universidad Aut\'{o}noma de Madrid, 28049 Madrid, Spain\\
$^{3}$Kobayashi-Maskawa Institute for the Origin of Particles and the Universe, Nagoya University, Chikusa-ku, Nagoya, Aichi, 464-8602, Japan
}

\date{Accepted XXX. Received YYY; in original form ZZZ}

\pubyear{\the\year{}}

\begin{document}

\label{firstpage}
\pagerange{\pageref{firstpage}--\pageref{lastpage}}
\maketitle

% Abstract of the paper
\begin{abstract}
Recent pulsar timing array (PTA) observations have reported evidence of a gravitational wave background (GWB). If supermassive black holes (SMBHs) are indeed the primary source of this signal, future PTA observations, such as those from the Square Kilometer Array (SKA), are expected to simultaneously capture multiple continuous gravitational waves (CGWs) emitted by bright individual SMBH binaries alongside a gravitational wave background (GWB). To address this anticipated scenario in the SKA era, we revisit the $\mathcal{F}$-statistic, a detection method for single-source signals in PTA datasets, and introduce a new model that accounts for unresolved GWs as a stochastic GWB. Here, we applied this improved $\mathcal{F}$-statistic to the datasets that include both CGW and GWB and evaluated how accurately the $\mathcal{F}$-statistic can identify the parameters of CGW. As a result, we demonstrate that our approach can successfully improve the estimation of the sky position and the amplitude of CGW, particularly when the GWB is dominant over white noise. 
This work serves as an initial step toward developing an efficient and robust algorithm based on the $\mathcal{F}$-statistic for future PTA observations.
\end{abstract}

\section{Introduction}
Since the direct detection of gravitational waves (GWs) by the LIGO collaboration in 2015 \citep{2016PhRvL.116f1102A}, GW observations have been increasingly recognised as a powerful tool for exploring new physics. Pulsar timing array (PTA) can probe GWs in the low-frequency range of nHz-$\mu$Hz by observing multiple millisecond pulsars (MSPs) over long periods and analysing the times of arrival of their radio pulses. Several PTA projects are currently operating worldwide, such as NANOGrav \citep{2013CQGra..30v4008M}, EPTA \citep{2013CQGra..30v4009K}, PPTA \citep{2013PASA...30...17M}, IPTA \citep{2010CQGra..27h4013H}, InPTA \citep{2018JApA...39...51J}, CPTA \citep{2016ASPC..502...19L}, and MPTA \citep{2023MNRAS.519.3976M}.
The next-generation PTA is planned to be developed with the Square Kilometre Array (SKA; \citealt{2015aska.confE..37J}). SKA is an ongoing project that aims to build approximately 200 parabolic radio dishes and 130,000 low-frequency dipoles, effectively forming a large telescope with a total collecting area exceeding one square kilometre. Currently, the first phase of the observational plan (SKA1-Mid) involves constructing parabolic antennas in South Africa. 
For the second phase (SKA2-Mid), projections suggest that around 6,000 MSPs will be detected in total, approximately 1,000 with noise levels below 100 ns, and several hundred with noise levels below 50 ns \citep{2020PASA...37....2W}. 
Therefore, future PTA observation in the SKA era will be able to achieve exceptional sensitivity and have the potential to probe GWs much more deeply than current PTAs.

One of the primary aims of PTA observations is to detect the superposition of GWs emitted by supermassive black holes (SMBHs). SMBHs typically have masses ranging from $10^{6}M_\odot$ to $10^{10}M_\odot$, and it is still uncertain how SMBHs have formed and evolved (see, e.g., \citealt{2013ARA&A..51..511K}, \citealt{2020ARA&A..58...27I}).
Theoretical predictions indicate that the GWB from the SMBH population reflects the evolutionary processes of SMBHs (see, e.g., \citealt{2004ApJ...615...19E}, \citealt{2023ApJ...959..117F}). Therefore, PTAs are expected to find important evidence regarding the evolution of SMBHs.
There are also other types of GW sources with cosmological origins, such as inflationary GWs and cosmic strings (see, e.g., \citealt{2023ApJ...951L..11A} and \citealt{2024A&A...685A..94E} for reviews). Thus, future PTA observations will be increasingly valuable and significant for advancing our understanding of modern cosmology.

Recently, PTA observation collaborations worldwide (NANOGrav; \citealt{2023ApJ...951L...8A}, EPTA and InPTA; \citealt{2023arXiv230616214A}, PPTA; \citealt{2023ApJ...951L...6R}, CPTA; \citealt{2023RAA....23g5024X}, MPTA; \citealt{2025MNRAS.536.1489M}) have reported evidence of GW background (GWB).
The primary candidate for the origin of this signal is SMBHs. 
Among the superposition of these GWs, those generated from nearby SMBH binaries can be bright enough to be resolved individually. 
Consequently, it is highly anticipated that multiple deterministic continuous gravitational waves (CGWs) and a stochastic GWB will likely be observed simultaneously in future PTA observations (see, e.g., \citealt{2015MNRAS.451.2417R} and \citealt{2018MNRAS.477..964K}).

\cite{2023ApJ...959....9B} and \cite{2024arXiv240721105F} performed the Bayesian parameter estimation for the injected GWB in the mock data simulation, considering specific SMBH scenarios. 
Their results clearly indicate that, if we analyse the data assuming only the existence of GWB, the presence of CGW causes bias in the predictions of the GWB spectrum.
Therefore, as we look toward the future of PTAs in the SKA era, the existence of CGW is becoming increasingly significant, and it is essential to develop a detection method that can efficiently identify multiple CGWs and GWB simultaneously (see e.g. \citealt{2022PhRvD.105l2003B}).

To deal with this anticipated scenario, in this study, we revisit the $\mathcal{F}$-statistic, a detection method for single-source signals, first introduced in \cite{1998PhRvD..58f3001J}. 
The $\mathcal{F}$-statistic estimates the parameters of injected CGW through a maximization of the likelihood. One notable advantage of the $\mathcal{F}$-statistic is that it reduces the number of parameters to be optimised through an analytical maximization of the likelihood. Previous works have applied the $\mathcal{F}$-statistic to PTA observation and evaluated its performance. 
\cite{2012PhRvD..85d4034B} performed a simulation of mock data for PTA observations under a simplified assumption and setup. They demonstrated that given a sufficient number of pulsars, the $\mathcal{F}$-statistic is capable of resolving multiple GW sources in PTA observations.
Based on this work, \cite{2013PhRvD..87f4036P} combined genetic algorithms with the $\mathcal{F}$-statistic to develop an efficient method for resolving multiple CGWs. 
Conversely, \cite{2012ApJ...756..175E} provided a more in-depth discussion of the methodology behind the $\mathcal{F}$-statistic, taking into account factors such as deterministic noises and the influence of pulsar terms. They also introduced a pipeline for detecting CGWs based on the $\mathcal{F}$-statistic.
Following these works, the pipeline based on the $\mathcal{F}$-statistic was then applied to the real observational datasets from EPTA and placed an upper limit on a CGW from an SMBH binary in \cite{2016MNRAS.455.1665B}.

In those previous works, the $\mathcal{F}$-statistic has consistently been applied under the assumption that the GWB component is subdominant relative to the noise. The validity of the $\mathcal{F}$-statistic for parameter estimation in the presence of a significant GWB has not yet been thoroughly examined.
In this study, to prepare for the expected conditions in the SKA era, we begin by assessing the performance of the $\mathcal{F}$-statistic in the presence of a GWB using mock data simulations. Then we present a new model where we include a contribution from unresolved GWs, namely a SGWB, in the noise matrix. 
We apply our improved $\mathcal{F}$-statistic to mock datasets that include both single CGW and SGWB in both simple and realistic scenarios based on an ideal SMBH population model. Finally, we discuss the effectiveness of this statistical approach in the SKA era.

The motivation of this paper is to examine the performance of the $\mathcal{F}$-statistic in relatively simple settings to gain detailed insights into its behaviour. 
We are particularly interested in the bias that arises with a non-negligible GWB contribution. Rather than performing a full Bayesian analysis as done in previous studies, we use frequentist evaluations to address the issue. Our approach reduces computation time and allows us to test multiple realizations with varying pulsar sky distributions. Due to the limited number of available pulsars, simulation results can vary significantly based on the relative locations of the pulsars and CGW sources. Running multiple realizations helps us assess such statistical fluctuations in the results. In addition, the $\mathcal{F}$-statistic is computationally feasible for multiple sources, as demonstrated by \cite{2012PhRvD..85d4034B}.
Therefore, exploring this approach could be highly beneficial for developing efficient analysis techniques in the SKA era, where identifying multiple sources will be crucial.

The structure of this paper is as follows. 
In Section~\ref{sec:our modelling}, we describe the basic setup of our mock data simulation and the methodology of the $\mathcal{F}$-statistic.
In Section~\ref{sec:results}, we show the results of our mock data simulation and evaluate the performance of $\mathcal{F}$-statistic in the presence of GWB.
In Section~\ref{sec:summary&discussion}, we summarise our study and discuss prospects to establish the algorithm based on $\mathcal{F}$-statistic toward future PTAs in the SKA era.
The calculations in this study adopt the flat $\Lambda\mathrm{CDM}$ model with $h = 0.7$, $\Omega_{\mathrm{m}}=0.3$ and  $\Omega_{\Lambda}=0.7$. 

\section{Simulation setup}\label{sec:our modelling}
In this section, we introduce the basic setup of our mock data simulations and the methodology of the $\mathcal{F}$-statistic.
Our goal is to demonstrate the potential of the $\mathcal{F}$-statistic in estimating the parameters of the injected CGW and to analyse its performance in the anticipated scenario in the SKA era.
Although realistic PTA observations involve complex systematics, as a first step, we make simple assumptions and construct the timing residuals using minimal components, including only GWs and white noise.
Our assumptions in mock data simulations follow the simplified framework adopted in \cite{2012PhRvD..85d4034B}, and we follow \cite{2012ApJ...756..175E} for our mathematical formulation.

\subsection{timing model}\label{subsec:timing model}
We prepare $N_\mathrm{p}$ pulsars and assume observation of the difference in timing of arrivals (TOAs) of their pulses $N_\mathrm{TOA}$ times over the total observation period $T_\mathrm{obs}$.
A timing residual for each pulsar $x_\alpha$ has then $N_\mathrm{TOA}$ data points and is described as an $N_\mathrm{TOA}$-dimensional vector. Our timing residual is expressed as a combination of both signal $s_\alpha$ and noise components $n_\alpha$, given as 
\begin{equation}
\label{our dataset}
    x_\alpha = s_\alpha + n_\alpha~,
\end{equation}
where $\alpha = 1,2,\cdots, N_\mathrm{p}$ is the index of pulsars. 

\subsubsection{signal component}
The signal component is assumed to be a CGW produced by a single SMBH binary.
In general, GWs are described by a tensor metric perturbation, which can be decomposed into two polarization modes
\begin{equation}
    h_{\mu\nu}(t,\hat{\Omega}) = \sum_{A=+, \times} h_{A}(t,\hat{\Omega})e_{\mu\nu}^A(\hat{\Omega})~,
\end{equation}
where $\mu, \nu = 0, 1, 2, 3$ are the indices of the coordinates, $\hat{\Omega}$ is the unit vector corresponding to the direction of GW propagation, and $e_{\mu\nu}^A(\hat{\Omega})$ are the polarization tensors for each polarization state $A=+,\times$. 
By defining two additional unit vectors $\hat{u}$ and $\hat{v}$ that are orthogonal to $\hat{\Omega}$, the polarization tensors can then be expressed as
\begin{equation}
\begin{split}
    e_{\mu\nu}^+ &= \hat{u}_\mu\hat{u}_\nu - \hat{v}_\mu\hat{v}_\nu~,\\
    e_{\mu\nu}^\times &= \hat{u}_\mu\hat{v}_\nu + \hat{u}_\mu\hat{v}_\nu \,.
\end{split}
\end{equation}
In the observer's coodinate system, $\hat{x}$, $\hat{y}$, $\hat{z}$, these three unit vectors $\hat{\Omega}, \hat{u}, \hat{v}$ are defined by 
\begin{equation}
\begin{split}
\hat{\Omega} &= -(\sin\theta\cos\phi)\hat{x}-(\sin\theta\sin\phi)\hat{y}-(\cos\theta)\hat{z},\\
\hat{u} &= (-\sin\phi)\hat{x}+(\cos\phi)\hat{y},\\
\hat{v} &= (-\cos\theta\cos\phi)\hat{x} + (-\cos\theta\sin\phi)\hat{y} + (\sin\theta)\hat{z}.
\end{split}
\end{equation}
Here, $\phi$ and $\theta$ represent the angular coordinates that specify the sky position of the injected GW source.

Assuming that an SMBH binary is inspiralling and has a circular orbit, we can describe the two polarization modes of GWs generated by the SMBH binary with its masses $M_1$ and $M_2$ at redshift $z$ as 
\begin{equation}
\label{GW_from_SMBHB}
\begin{split}
h_+(t) & = 4\frac{(G\mathcal{M}_c)^{5/3}}{c^4d_L}(\pi f)^{2/3}\frac{1+\cos^2\iota}{2}\cos2\Phi(t)~,\\
h_\times(t) &= 4\frac{(G\mathcal{M}_c)^{5/3}}{c^4d_L}(\pi f)^{2/3}\cos\iota\sin2\Phi(t) \,.
\end{split}
\end{equation}
Here $\mathcal{M}_c = (1+z)M_c$ is a redshifted chirp mass with the chirp mass being $M_c=(M_1M_2)^{3/5}/(M_1+M_2)^{1/5}$, $\iota$ is an inclination of a binary system, and $f$ is the GW frequency in the observer's system which is related to the orbital frequency via $\omega = \pi f$.
If the change in the orbital frequency $\dot{f}$ is negligible during the observational period, we can approximate the GW phase $\Phi(t)$ as $\Phi(t) = \omega t + \Phi_0$, where $\Phi_0$ is the initial phase. 
Finally, $d_L$ denotes the luminosity distance to an SMBH binary, which can be calculated using as
\begin{equation}
    d_L = (1+z)\frac{c}{H_0}\int_{0}^z\frac{\dd z^\prime}{\sqrt{\Omega_\mathrm{m}(1+z^\prime)^3+\Omega_\Lambda}}~,\\
\end{equation}
where $H_0=100 h~ {\rm km}/{\rm sec}/{\rm Mpc}$ is the Hubble constant.

When considering the polarization angle of the binary system $\psi$, we need to transform two modes of GW as follows,
\begin{equation}
\label{psi_rotation_of_GW}
\begin{split}
h_+^\prime(t) &= h_+(t)\cos2\psi - h_\times \sin2\psi~,\\
h_\times^\prime(t) &= h_\times(t)\cos2\psi + h_+ \sin2\psi \,.\\
\end{split}
\end{equation}
By averaging over the inclination and polarization angles of inspiralling SMBH binaries in circular orbits, we obtain the average GW strain amplitude
\begin{equation}
\label{single GW strain}
h_{s}(f) = \frac{8}{10^{1/2}}\frac{(G\mathcal{M}_c)^{5/3}}{c^4d_L}(\pi f)^{2/3}.
\end{equation}
The characteristic strain amplitude can be obtained by taking into account the number of cycles in one frequency bin $\Delta f = 1/T_\mathrm{obs}$ during the observational period $T_\mathrm{obs}$ as, 
\begin{equation}
h_{\mathrm{CGW}, c}(f) = h_s(f)\sqrt{f/\Delta f} \,.
\end{equation}

Let us represent $t$ as the observer's time on the earth and $t_{\mathrm{p},\alpha}$ as the time when GW passes pulsar-$\alpha$. The difference in $h_A$ between the earth and the pulsar-$\alpha$ is expressed as $\Delta h_A(t, \hat{\Omega}) = h_A(t_{\mathrm{p},\alpha},\hat{\Omega})- h_A(t, \hat{\Omega})$. Then the timing residual induced by GWs consists of two contributions, the Earth term $s_\alpha^\mathrm{E}$ and the pulsar term $s_\alpha^\mathrm{P}$:
\begin{equation}
\label{timing_residual_GW}
\begin{split}
    s_\alpha(t) = &\sum_{A=+,\times}F_\alpha^{A}(\hat{\Omega})\int_{0}^{t}\dd t^\prime \Delta h_A(t^\prime,\hat{\Omega})\\
    &= s^\mathrm{P}_\alpha(t_{\mathrm{p},\alpha}) - s^\mathrm{E}_\alpha(t) \,.
\end{split}
\end{equation}
Here, $F^A_\alpha(\hat{\Omega})$ is the antenna pattern for each polarization, whose mathematical expression is given by
\begin{equation}
    F^A_\alpha(\hat{\Omega}) = \frac{1}{2}\frac{\hat{p}_\alpha^\mu\hat{p}_\alpha^\nu}{1+\hat{\Omega}\cdot\hat{p}_\alpha}e^A_{\mu\nu}(\hat{\Omega})~,
\end{equation}
where $\hat{p}_\alpha$ is the unit vector that points toward the sky position of pulsar-$\alpha$.

In this study, we construct the timing residual without considering the contribution from the pulsar term for the following reason. The frequency and phase of the pulsar term for pulsar-$\alpha$, $f_{\mathrm{p},\alpha}$ and $\Phi_{\mathrm{p}, \alpha}, $ depend on the distance to the pulsar, $L_\mathrm{p,\alpha}$, given as, 
\begin{equation}
\begin{split}
    f_{\mathrm{p},\alpha} &= f\Big[1-\frac{256}{5}\Big(\frac{G\mathcal{M}_c}{c^2}\Big)^{5/3}(\pi f)^{8/3}t_{\mathrm{p}, \alpha}\Big]^{-3/8}~,\\
    \Phi_\mathrm{p,\alpha} &= \pi f_{\mathrm{p},\alpha}t_{\mathrm{p},\alpha} + \Phi_0~,
\end{split}
\end{equation}
where $t_{\mathrm{p},\alpha} = t - (1+\hat{\Omega}\cdot\hat{p})L_{\mathrm{p},\alpha}/c$.
In cases where the Earth term has a frequency $f \gtrsim 10$~nHz, which is the scenario examined in this work, the pulsar term typically has a lower frequency compared to the Earth term and does not affect the analysis focusing on the observed frequency $f_\mathrm{obs} \gtrsim 10$~nHz. This assumption justifies neglecting the pulsar term, as supported by \cite{2012PhRvD..85d4034B}. Even if pulsar terms influence the observed frequency, the distances to individual pulsars are effectively random, causing the pulsar terms to sum incoherently and be suppressed, while summing the timing residuals enhances the coherent Earth term. 
Although neglecting pulsar terms can cause bias in the parameter estimation of GWs (see e.g., \citealt{2016MNRAS.461.1317Z} and \citealt{2024MNRAS.535..132K}),
\cite{2012ApJ...756..175E} suggests that, through this averaging process, the contribution from the pulsar term becomes negligible when the number of pulsars exceeds 50. Since the SKA era is expected to provide observations of potentially over one hundred pulsars, neglecting the pulsar term remains valid in this context.

For the reasons above, the timing residual coming from a single SMBH binary is described only by the Earth term. Therefore, we calculate the induced timing residual as
\begin{equation}
    s_\alpha(t) = -s^\mathrm{E}_\alpha(t) = F^+_\alpha(\hat{\Omega}) s_+(t) + F^\times_\alpha(\hat{\Omega}) s_\times(t).
\end{equation}
Here, the explicit forms of $s_+$, $s_\times$ are derived by substituting Eqs.~\eqref{GW_from_SMBHB} and \eqref{psi_rotation_of_GW} into Eq.~\eqref{timing_residual_GW}, 
\begin{equation}
\label{CGW_explicit}
\begin{split}
    s_+(t) = 2\frac{(G\mathcal{M}_c)^{5/3}}{c^3d_L}&(\pi f)^{-1/3}\Big[\sin2\Phi(t)\frac{1+\cos^2\iota}{2}\cos2\psi\\
    &+ \cos2\Phi(t)\cos\iota\sin2\psi\Big]\\
    s_\times(t) = 2\frac{(G\mathcal{M}_c)^{5/3}}{c^3d_L}&(\pi f)^{-1/3}\Big[\sin2\Phi(t)\frac{1+\cos^2\iota}{2}\sin2\psi \\
    &- \cos2\Phi(t)\cos\iota\cos2\psi\Big].
\end{split}
\end{equation}

\subsubsection{noise component}
\label{sec:noise}
The key distinction between our modelling and previous works (\citealt{2012PhRvD..85d4034B}, \citealt{2012ApJ...756..175E}) is that we account for unresolved GWs as an additional noise component. With the expected high sensitivity of future PTAs, the data will contain two types of GW signals: (i) CGWs, which are bright enough to be resolved individually, and (ii) the GWB, which represents an ensemble of unresolved GWs. The CGWs are deterministic and can be described analytically, while the GWB is stochastic. In our approach, we treat the stochastic GWB as a noise component when applying the $\mathcal{F}$-statistics to the search for the CCWs.

Therefore, the noise component is modelled as a combination of two contributions: white noise $n_{\mathrm{WN}, \alpha}$ and the GWB $n_{\mathrm{GWB}, \alpha}$,
\begin{equation}
    n_\alpha = n_\mathrm{\mathrm{WN}, \alpha}+ n_{\mathrm{GWB}, \alpha} \,.
\end{equation}
The white noise is stationary and Gaussian, with a mean of zero and a root mean square of $\sigma_{\alpha}$.
The GWB is the ensemble of unresolved GWs from all directions in the sky, which we model as follows:
First, to characterise the stochastic GWB, we decompose the metric perturbations into Fourier components,
\begin{equation}
    h_{\mu\nu}(t, \vec{r}) = \sum_{A=+, \times}\int_{-\infty}^{\infty}\dd f \int \dd\hat{\Omega} \tilde{h}_{A}(f, \hat{\Omega})e^A_{\mu\nu}(\hat{\Omega})\ee^{2\pi if(t-\hat{\Omega}\cdot\vec{r})} \,.
\end{equation}
We assume that GWB is isotropic, unpolarised, and Gaussian. Then, it is fully characterised by the cross-correlation of its Fourier components, given by 
\begin{equation}
    \langle h^*_A(f, \hat{\Omega})h_{A^\prime}(f^\prime,\hat{\Omega}^{\prime}) \rangle = \frac{S(f)}{16\pi}\delta(f-f^\prime)\delta^2({\hat{\Omega},\hat{\Omega}^\prime})\delta_{A,A^\prime}~,
\end{equation}
where $S(f)$ is the power spectral density.
The characteristic strain is defined as $h_{\mathrm{GWB}, c}(f) \equiv \sqrt{fS(f)}$. Then we assume that the spectral amplitude of the GWB follows a power-law spectrum, expressed as
\begin{equation}
    h_{\mathrm{GWB}, c}(f) =  A_g\Big(\frac{f}{f_\mathrm{ref}}\Big)^{\alpha_g} \,.
    \label{eq:power_law}
\end{equation}
Especially if we assume all SMBH binaries have circular orbits, the power-law index of GWB generated from SMBH binaries is $\alpha_g = -2/3$ as explained in Section~\ref{sec:results}. According to the latest observational results from NANOGrav \citep{2023ApJ...951L...8A}, the estimated amplitude of the GWB is $A_g=2.4_{-0.6}^{+0.7}\times 10^{-15}$, assuming that the observed GWB originates from SMBHs with $\alpha_g=-2/3$.

Based on these assumptions, we simulate the timing residual induced by a GWB, following \cite{2009MNRAS.394.1945H}. We consider $N_\mathrm{GWB}$ monotonic GWs to construct a GWB and assume the number of the GW modes $N_\mathrm{mode}$. The minimum and maximum frequency are given by $f_\mathrm{L}=1/T_\mathrm{obs}$ and $f_\mathrm{H}=N_\mathrm{mode}/T_\mathrm{obs}$, respectively. We generate $N_\mathrm{GWB}$ random numbers which are uniformly distributed between $f_\mathrm{L}-\frac{1}{2}\Delta f$ and $f_\mathrm{H}+\frac{1}{2}\Delta f$ in log space. We determine the number of GW sources for the $i$-th frequency $n_i$ equal to the number of random numbers in the frequency range [$f_i-\frac{1}{2}\Delta f$, $f_i+\frac{1}{2}\Delta f$]. The directions of the injected GW sources are selected uniformly across the sky. Using Eq.~\eqref{GW_from_SMBHB}, we then simulate the timing residual induced by the GWB, resulting from the ensemble of these multiple GWs as
\begin{equation}
    n_\mathrm{GWB,\alpha} = \sum_{i=1}^{N_\mathrm{mode}}\sum_{j=1}^{n_i}(F_\alpha^+r_{+, j} + F_\alpha^{\times}r_{\times, j}) \,.
\end{equation}
Here, we neglect the contributions from the pulsar terms. Each $r_{A, j}$ is calculated by
\begin{equation}
    r_{A,j}(t) = \tilde{a}_{A, j}\sin\Big(\frac{2\pi i}{T_\mathrm{obs}}t\Big) + \tilde{b}_{A, j}\cos\Big(\frac{2\pi i}{T_\mathrm{obs}}t\Big) \,.
\end{equation}
The amplitudes of its Fourier components $\tilde{a}_{A,j}$ and $\tilde{b}_{A,j}$ are determined by following a Gaussian distribution with zero mean and the variance of $\sigma_{\mathrm{GWB}, j}$, where  
\begin{equation}
\sigma_{\mathrm{GWB},j} = \sqrt{\frac{\ln(f_\mathrm{H}/f_\mathrm{L})}{N_\mathrm{GWB}}\frac{h_{\mathrm{GWB},c}^2}{16\pi^2f_i^2}} \,.
\end{equation} 
In this work, we take $N_\mathrm{mode}=30$ and $N_\mathrm{GWB}=10^4$.

\subsection{Formulation of the Likelihood}\label{subsec:likelihood}
We now introduce the likelihood function, which will be used in the $\mathcal{F}$-statistic described in the next subsection.
Hereafter, we refer to the series of timing residuals of each pulsar as a vector with $N_\mathrm{TOA}$ components.
When written in bold, the timing residual represents an $N_\mathrm{TOA}N_\mathrm{p}$ vector, where the residuals of all pulsars are stacked vertically as follows:
\begin{equation}
\label{bold notation}
\vb{x}=\mqty(
x_1\\
x_2\\
\vdots\\
x_{N_\mathrm{p}}
), 
\vb{s}=\mqty(
s_1\\
s_2\\
\vdots\\
s_{N_\mathrm{p}}
), 
\vb{n}=\mqty(
n_1\\
n_2\\
\vdots\\
n_{N_\mathrm{p}}
)~,
\end{equation}
where $x_\alpha$, $s_\alpha$ and $n_\alpha$ is the data vector of pulsar-$\alpha$ with $N_{\rm TOA}$ components defined in Eq.~\eqref{our dataset}.

Since the noise component is stationary and Gaussian, its likelihood $\mathcal{L}$ should be expressed as
\begin{equation}
\label{eq:likelihood}
    \mathcal{L}(\vb{n}) = \frac{1}{\sqrt{\det2\pi\Sigma_\mathrm{n}}}\exp\Big(-\frac{1}{2}\vb{n}^T\Sigma_\mathrm{n}^{-1}\vb{n}\Big) \,.
\end{equation}
Here $\Sigma_\mathrm{n}$ is a noise matrix, and this quantity reflects the modelling of the noise component, given by
\begin{equation}
    \Sigma_\mathrm{n} = \langle \vb{n}\vb{n}^T \rangle,
\end{equation}
where the notation $\langle\cdots\rangle$ expresses an ensemble average of the quantity inside. The noise matrix is composed of two components: white noise and the GWB. 
The white noise is an intrinsic noise of each pulsar consisting of the pulse phase jitter noise and the radiometer noise; thus, there is no correlation between the white noises of all pulsars.
\begin{equation}
    \Sigma_\mathrm{WN,\alpha\beta} = \langle\vb{n}_{\mathrm{WN},\alpha}\vb{n}_{\mathrm{WN},\beta}^T\rangle = \sigma^2_\alpha\delta_{\alpha\beta}I_{N_\mathrm{TOA}}, 
\end{equation}
where $I_{N_\mathrm{TOA}}$ is a $N_\mathrm{TOA}$ identity matrix. 
On the other hand, GWB induces a common signal in the timing residuals of all pulsars, and thus, there is a correlation between the timing residuals of all pulsars. The contribution of an isotropic GWB to the noise matrix is given by, 
\begin{equation}
\Sigma_{\mathrm{GWB},\alpha\beta} = \langle \vb{n}_{\mathrm{GWB},\alpha} \vb{n}_{\mathrm{GWB}, \beta}^{T} \rangle= \zeta_{\alpha\beta}F\varphi_{\mathrm{GWB}} F^{T}.
\label{noise matrix of GWB}
\end{equation}
Here, $\zeta_{\alpha\beta}$ is the spatial correlation called Hellings \& Downs curve, which uniquely appears in the existence of an isotropic GWB \citep{1983ApJ...265L..39H}. 
This spatial correlation can be derived from the cross-correlation between antenna patterns of the pulsar pair, calculated as,  
\begin{equation}
\begin{split}
    \zeta_{\alpha\beta} &= \frac{3}{2}\int\frac{\dd \hat{\Omega}}{4\pi}\sum_{A=+, \times} F_\alpha^A(\hat{\Omega}) F_\beta^A(\hat{\Omega})\\
    &= \frac{3}{2}\chi_{\alpha\beta}\ln{\chi_{\alpha\beta}}-\frac{1}{4}\chi_{\alpha\beta}+\frac{1}{2}.
\end{split}
\end{equation}
where we define $\chi_{\alpha\beta} = (1 - \cos\theta_{\alpha\beta})/2$ with $\theta_{\alpha\beta}$ denoting the angle between pulsars $\alpha$ and $\beta$.
Note that, since we also ignore the contribution of pulsar terms in this simulation, $\zeta_{\alpha\beta}\neq 1$ in the case $\alpha=\beta$. 
$F$ in Eq.~(\ref{noise matrix of GWB}) is the Fourier design matrix with dimension $N_\mathrm{TOA} \times 2N_\mathrm{mode}$. For $j = 1, \dots, 2N_\mathrm{mode}$, the $(2j-1)$-th column corresponds to the Fourier basis $\sin(2\pi f_j t)$, whereas the $2j$-th column corresponds to $\cos(2\pi f_j t)$. The covariance matrix between the Fourier components $\varphi_\mathrm{GWB}$ has the $2N_\mathrm{mode}\times2N_\mathrm{mode}$ dimension. Since we assume GWB follows stationarity and Gaussian,  this matrix is diagonal, with each component given by the power spectral density for each frequency bin, given as,
\begin{equation}
\varphi_{\mathrm{GWB}} = \frac{1}{T_\mathrm{obs}}\mqty(\frac{h_c^2(f_1)}{12\pi^2f_1^3} & 0 & \cdots & 0\\
0 & \frac{h_c^2(f_1)}{12\pi^2f_1^3} & \cdots & 0\\
\vdots & \vdots & \ddots & \vdots\\
0 & 0 & \cdots & \frac{h_c^2(f_{N_\mathrm{mode}})}{12\pi^2f_{N_\mathrm{mode}}^3}
) \,.
\end{equation}

\subsection{$\mathcal{F}$-statistic}\label{subsec:F-statistic}
In this subsection, we introduce the methodology of the $\mathcal{F}$-statistic.\footnote[1]{As mentioned in Section~\ref{sec:our modelling}, we neglect the contribution of the pulsar term. Therefore, our formulation here corresponds to $\mathcal{F}_e$-statistic in \cite{2012ApJ...756..175E}.
}
The $\mathcal{F}$-statistic is a detection method for identifying single-source sinusoidal wave signals by maximizing the likelihood function.
First, the log-likelihood is analytically maximised with respect to the so-called extrinsic parameters.
Then, by numerically maximizing it over the intrinsic parameters, we can reconstruct CGW signals.
When applying the $\mathcal{F}$-statistic, we fix the observational frequency and predefine the number of CGW sources, $N_\mathrm{CGW}$, in the dataset.

Hereafter, we describe the procedure of the $\mathcal{F}$-statistic for the case of a single CGW source, i.e., $N_\mathrm{CGW} = 1$.
We define the inner product of two datasets $\vb{x}$ and $\vb{y}$ with a noise matrix $\Sigma_\mathrm{n}$ as, 
\begin{equation}
    \langle\vb{x}|\vb{y}\rangle \equiv \vb{x}^{T}\Sigma_\mathrm{n}^{-1}\vb{y} \,.
\end{equation}
As described in Eq.~\eqref{CGW_explicit}, the single-source signal is described by a total of six parameters. The parameters for the source position ($\phi$, $\theta$) are treated as the intrinsic parameters, while ($\xi=(GM_c)^{5/3}/(c^3d_L)$, $\iota$, $\psi$, $\Phi_0$) are treated as the extrinsic parameters. By explicitly separating the extrinsic and intrinsic parameters, the CGW signal can be expressed as:
\begin{equation}
\label{CGW_temp}
    s_\alpha = \sum_{i=1}^{4}a_i(\xi,\iota,\psi,\Phi_0)A^{i}_{\alpha}(t,\phi,\theta)~,
\end{equation}
where the explicit forms of extrinsic and intrinsic factors, $a_i(\xi,\iota, \psi, \Phi_0)$ and $A_\alpha^i(\phi, \theta, t)$ are expressed as,
\begin{equation}
\label{ai_explicit}
\begin{split}
a_{1} = \xi[(1+\cos^2\iota)&\cos2\Phi_0\cos2\psi\\
&-2\cos\iota\sin2\Phi_0\sin2\psi]\\
a_{2} = \xi[(1+\cos^2\iota)&\sin2\Phi_0\cos2\psi\\
&+2\cos\iota\cos2\Phi_0\sin2\psi]\\
a_{3} = \xi[(1+\cos^2\iota)&\cos2\Phi_0\sin2\psi\\
&+2\cos\iota\sin2\Phi_0\cos2\psi]\\
a_{4} = \xi[(1+\cos^2\iota)&\sin2\Phi_0\sin2\psi\\
&-2\cos\iota\cos2\Phi_0\cos2\psi],\\
\end{split}
\end{equation}
and,
\begin{equation}
\label{Ai_explicit}
\begin{split}
A_\alpha^1 = F_\alpha^+(\hat{\Omega})(\pi f_\mathrm{obs})^{-1/3}\sin2\pi f_\mathrm{obs} t\\
A_\alpha^2 = F_\alpha^+(\hat{\Omega})(\pi f_\mathrm{obs})^{-1/3}\cos2\pi f_\mathrm{obs} t\\
A_\alpha^3 = F_\alpha^\times(\hat{\Omega})(\pi f_\mathrm{obs})^{-1/3}\sin2\pi f_\mathrm{obs} t\\
A_\alpha^4 = F_\alpha^\times(\hat{\Omega})(\pi f_\mathrm{obs})^{-1/3}\cos2\pi f_\mathrm{obs} t.\\
\end{split}
\end{equation}
When estimating the parameters of multiple CGW signals, the $\mathcal{F}$-statistic can be extended by incorporating additional $N_\mathrm{CGW}$ signal templates with the same form as Eq.~\eqref{CGW_temp}.
In this case, the index $i$ runs from $1$ to $4N_\mathrm{CGW}$, accounting for all sources.

In our analysis, the null hypothesis $\mathcal{H}_0$ assumes that the dataset contains only noise ($\vb{x} = \vb{n}$), while the alternative hypothesis $\mathcal{H}_1$ assumes the presence of a signal ($\vb{x} = \vb{s} + \vb{n}$). In this context, the likelihoods of these hypotheses are represented as $\mathcal{L}(\mathcal{H}_0)\propto e^{-\langle \vb{x}|\vb{x}\rangle}$ and $\mathcal{L}(\mathcal{H}_1)\propto e^{-\langle \vb{x-\vb{s}}|\vb{x}-\vb{s}\rangle}$ following Eq.~\eqref{eq:likelihood}. Therefore, the log-likelihood ratio $\ln \Lambda$ between these two hypotheses is given by
\begin{equation}
\begin{split}
\label{log-likelihood}
    \ln\Lambda &\equiv \ln\frac{\mathcal{L}(\vb{x}|\mathcal{H}_1)}{\mathcal{L}(\vb{x}|\mathcal{H}_0)} = \langle\vb{x}|\vb{s}\rangle-\frac{1}{2}\langle\vb{s}|\vb{s}\rangle\\ 
    &= a_iX^i-\frac{1}{2}a_iM^{ij}a_j~,\\
\end{split}
\end{equation}
where $X^i\equiv \langle \vb{x}|\vb{A}^i \rangle$ and $M^{ij}\equiv\langle \vb{A}^i|\vb{A}^j\rangle$ \footnote[2]{Here $\vb{A}^i$ denotes a $N_\mathrm{TOA}N_\mathrm{p}$ vector consisting of the timing residuals $A^i_\alpha$ of all pulsars ($\alpha=1,2,...,N_\mathrm{p}$) as Eq.~\eqref{bold notation}.}.
Since $\ln\Lambda$ is expressed in a quadratic form with respect to the extrinsic factors $a_i$, it can be analytically maximised over $a_i$ using the following procedure:
\begin{equation}
\begin{split}
    0 = &\pdv{\ln\Lambda}{a_k}\Big|_{a=\hat{a}} = X^k - M^{ik}\hat{a}_i\\
    &\therefore \hat{a}_i = M_{ij}X^j~,
\end{split}
\end{equation}
where the hat denotes the value at the maximum.
Using the optimised extrinsic parameters $\hat{a}_i$, the statistical quantity $\mathcal{F}$ is defined as the reduced log-likelihood ratio:
\begin{equation}
\label{reduced likelihood}
\mathcal{F}\equiv\ln\Lambda(a=\hat{a},\phi,\theta) = \frac{1}{2}X^iM_{ij}X^j \,.
\end{equation}
By numerically maximizing the statistical quantity $\mathcal{F}$, we can estimate the sky position of a CGW source, $(\hat{\phi}, \hat{\theta})$.
Once we have the values of $\hat{\phi}$ and $\hat{\theta}$, we can determine $\vb{A}^i$, $X^i$, and $M_{ij}$, which then allows us to compute the optimal extrinsic parameters $\hat{a}$.
Combining this $\hat{a}_i$ with the obtained $\hat{A}^i$, we can finally reproduce the CGW signal in the dataset.
Additionally, defining the supplementary quantities in terms of $\hat{a}$ as
\begin{equation}
\begin{split}
A_+ &=\sqrt{(\hat{a}_1+\hat{a}_4)^2+(\hat{a}_2-\hat{a}_3)^2} \\
&+ \sqrt{(\hat{a}_1-\hat{a}_4)^2+(\hat{a}_2+\hat{a}_3)^2}~,\\
A_\times &=\sqrt{(\hat{a}_1+\hat{a}_4)^2+(\hat{a}_2-\hat{a}_3)^2} \\
&- \sqrt{(\hat{a}_1-\hat{a}_4)^2+(\hat{a}_2+\hat{a}_3)^2}~,\\
A &= A_+ + \sqrt{A_+^2 + A_\times^2}~,
\end{split}
\end{equation}
the extrinsic parameters can be derived from $\hat{a}$ as follows:
\begin{equation}
\begin{split}
\iota &= \cos^{-1}\Big(-\frac{A_\times}{A}\Big)~,\\ 
\psi &= \frac{1}{2}\tan^{-1}\Big(\frac{A_+\hat{a}_4-A_\times \hat{a}_1}{A_+\hat{a}_2+A_\times \hat{a}_3}\Big)~,\\
\Phi_0 &= \frac{1}{2}\tan^{-1}\Big(\frac{A_+\hat{a}_2 + A_\times \hat{a}_3}{A_+\hat{a}_1 - A_\times \hat{a}_4}\Big)~,\\
\xi &= \frac{|A|}{4} \,.
\end{split}
\end{equation}

Note that $\mathcal{F}$ follows a non-central $\chi^2$ distribution with four degrees of freedom, and its non-centrality parameter is related to the square of signal-noise ratio, defined as $\mathrm{SNR}\equiv\sqrt{\langle \vb{s}|\vb{s} \rangle}$. 
Therefore, the expectation value of $\mathcal{F}$ is related to the SNR by the following relation
\begin{equation}
\label{expect value of F}
    E(\mathcal{F}) = \frac{1}{2}[\langle \vb{s}|\vb{s} \rangle + 4] \propto h_s^2(f_\mathrm{obs}) \,.
\end{equation}

\section{Evaluating performances of $\mathcal{F}$-statistic}\label{sec:results}
As mentioned in section~\ref{sec:noise}, the key difference in modelling compared to previous works lies in the noise matrix. 
The previous $\mathcal{F}$-statistic only accounted for white noise (as $\Sigma_\mathrm{n} = \Sigma_\mathrm{WN}$), whereas our modified $\mathcal{F}$-statistic includes an additional contribution from the GWB (as $\Sigma_\mathrm{n} = \Sigma_\mathrm{WN} + \Sigma_\mathrm{GWB}$).
We will refer to these two noise modelling approaches as the standard and the GWB-informed $\mathcal{F}$-statistic, respectively.
In this section, we examine how this difference in noise modelling impacts the behaviour of the statistical quantity $\mathcal{F}$ and the accuracy of CGW parameter estimation through mock data simulations.
In Section~\ref{subsec:simple cases} and Appendix~\ref{sec:appendix.A}, we construct test data composed of CGW, GWB, and white noise using simplified settings. By varying the parameters of each component, we statistically evaluate the performance of the $\mathcal{F}$-statistic in several cases.
In Section~\ref{subsec:realistic cases}, we consider more realistic scenarios by adopting the phenomenological SMBH population model, known as the Agnostic model \citep{2016MNRAS.455L..72M}, to construct the GW components and perform the $\mathcal{F}$-statistic analysis.

In the following simulations, the observation time is set to $T_\mathrm{obs} = 10$ $\mathrm{yr}$, and the cadence $\Delta t$ is set to $\Delta t = 2$ $\mathrm{weeks}$ for all pulsars.
The observational frequency $f_\mathrm{obs}$ is taken to be $f_\mathrm{obs} = 5/T_\mathrm{obs} = 16$ $\mathrm{nHz}$.
When performing the $\mathcal{F}$-statistic, we consider a situation where a single CGW is injected into the dataset ($N_\mathrm{CGW} = 1$) at $f=f_\mathrm{obs}$ for simplicity.
We calculate $\mathcal{F}$ at each point on the celestial sphere uniformly divided following \texttt{HEALPix} map \citep{2005ApJ...622..759G}. We adopt the Python library \texttt{healpy} \citep{Zonca2019} to generate \texttt{HEALPix} map. and then find its global maximum. 
We also assume information about the injected noise components (white noise and GWB) is known before applying our $\mathcal{F}$-statistic. 
This assumption is discussed further in Section~\ref{sec:summary&discussion} and Appendix~\ref{subsec:appendix.A3}. 
In a realistic pipeline, we envisage a Bayesian analysis in which the CGW parameters and the GWB parameters are inferred jointly.
The performance of $\mathcal{F}$-statistic is statistically evaluated by generating 100 realizations. 

\subsection{Simple case}\label{subsec:simple cases}
In this subsection, we demonstrate the mock data simulations using simplified settings, focusing on the case of $N_\mathrm{p}=20$ pulsars, all with an equal white noise level of $\sigma$. First, in Case 1, we compare the performance of the standard $\mathcal{F}$-statistic with and without GWB to illustrate how GWB affects CGW detection. Then, in Case 2, we compare the performance of two versions of $\mathcal{F}$-statistics to show how the GWB-informed modelling enhances performance. 

For the injection of a single CGW, we randomly vary the sky position while keeping the other parameters the same: $h_s(f_\mathrm{obs}) \simeq 1.1\times 10^{-15}$, $\iota = \pi/2$, $\psi = 0$, $\Phi_0 = 0$. 
When injecting the GWB, we assume the power-law spectrum, as described in Eq.~\eqref{eq:power_law}, with the amplitude $A_g = 1.0\times10^{-15}$ and the tilt $\alpha_g = -2/3$.
With this setting, the amplitude of a single CGW satisfies $\lambda_\mathrm{fore} \equiv h_{\mathrm{CGW},c}^2/(h_{\mathrm{CGW},c}^2 + h_{\mathrm{GWB},c}^2) = 0.7$. In each realization, we change the sampling of white noise and GWB, as well as the positions of pulsars and the direction of the injected CGW. 

\subsubsection{Case 1. Impact of the GWB presence}
First, we investigate how the presence of the GWB affects the performance of the standard $\mathcal{F}$-statistic, which considers only white noise in the noise matrix. We prepare two mock datasets: one without GWB and the other with GWB, characterised by $A_g=1.0\times10^{-15}$ and $\alpha_g=-2/3$. We then apply the $\mathcal{F}$-statistic to both datasets.

\begin{figure*}
\centering
\begin{minipage}[b]{\columnwidth}
    \centering
    \includegraphics[width=0.99\columnwidth]{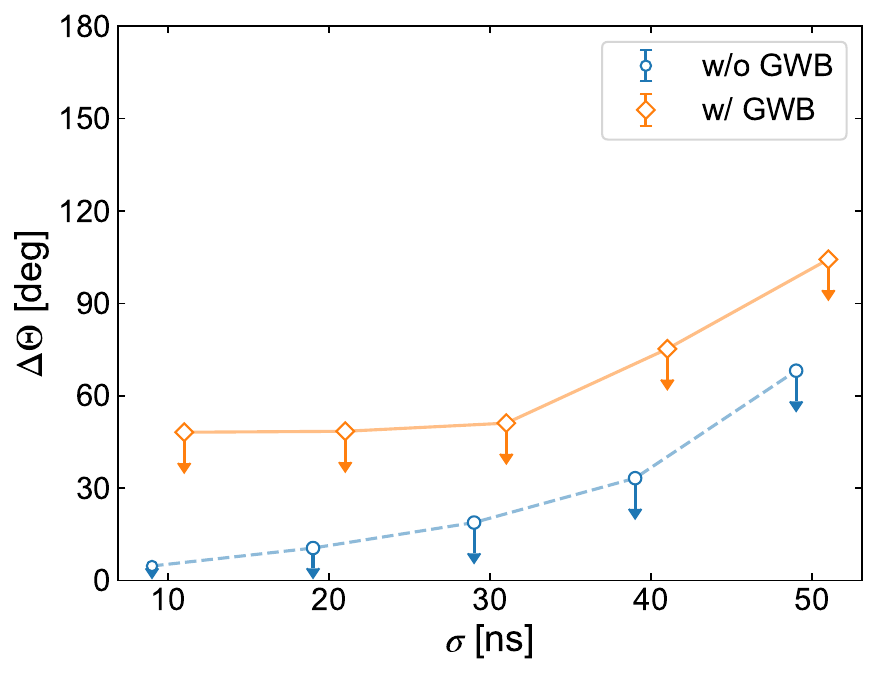}
\end{minipage}
\begin{minipage}[b]{\columnwidth}
    \centering
    \includegraphics[width=0.99\columnwidth]{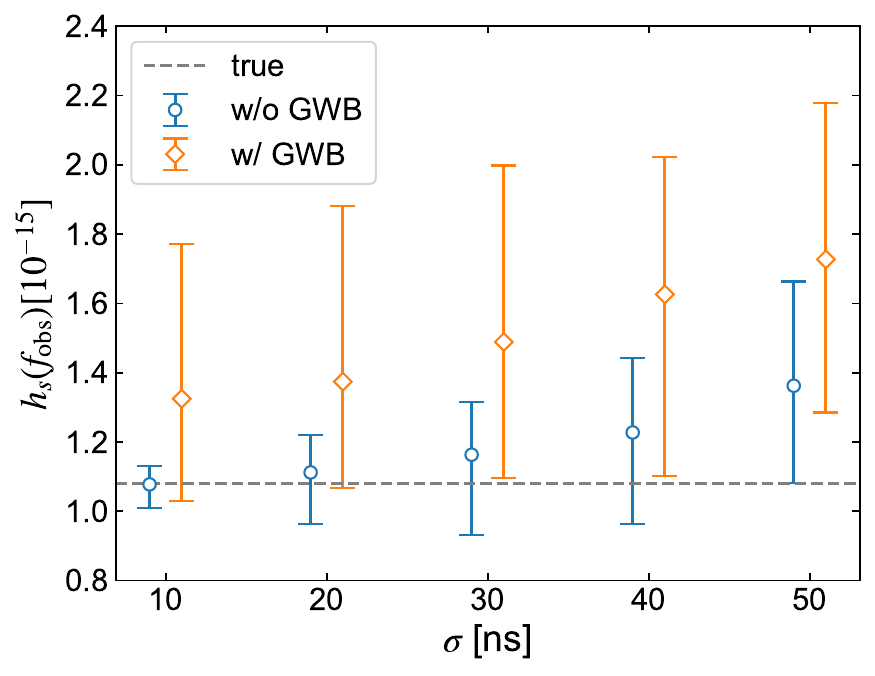}
\end{minipage}
\caption{Results of \textit{case 1}. Predicted parameter values obtained from 100 realizations by applying the standard $\mathcal{F}$-statistic to mock datasets: one dataset includes white noise and a single CGW (blue circles with error bars or downward arrows), while the other includes white noise, a single CGW, and GWB (red diamonds with error bars or downward arrows). The left panel illustrates the deviation of the source position from the true position of the injected CGW source, evaluated at different noise levels.
Each point with a downward arrow means that the deviation of the angle is within the indicated value for $68\%$ of the realizations.
The right panel shows the predicted GW strain amplitude of the injected CGW source for different noise levels. Each point and error bar shows the median and the $16\%$-$84\%$ region of each distribution, respectively. The gray dashed line represents the true GW strain amplitude of the injected signal.}
\label{fig:case1}
\end{figure*}

In Figure~\ref{fig:case1}, we show the distribution of the estimated parameter values of the injected CGW for different white noise levels. 
The left panel illustrates the distribution of $\Delta\Theta = \arccos(\hat{n}_\mathrm{true}\cdot\hat{n}_\mathrm{pred})$, where $\hat{n}_\mathrm{true}$ and $\hat{n}_\mathrm{pred}$ are the unit vectors representing the true and predicted directions of the CGW source, respectively. 
In the case of strong white noise, the presence of GWB does not significantly impact the accuracy of the sky position estimation for the CGW source. However, when GWB dominates over the white noise, the results indicate that the standard $\mathcal{F}$-statistic fails to accurately determine the direction of the injected CGW.

The right panel illustrates the predictions of the GW strain. It is clear that the presence of the GWB introduces additional uncertainty. Furthermore, we find that the standard $\mathcal{F}$-statistic tends to overestimate the CGW strain in the presence of a GWB. This overestimation arises because the modelling used in the standard $\mathcal{F}$-statistic misinterprets unresolved GWs as part of a single CGW signal during parameter estimation.

Furthermore, regardless of the presence of the GWB, there is a tendency for strain predictions to be overestimated as the white noise becomes stronger. 
This tendency can be qualitatively explained by considering the contribution of noise components in $\mathcal{F}$.
Defining $N^i\equiv\langle\vb{n}|\vb{A}^{i} \rangle$ and $S^i\equiv\langle\vb{s}|\vb{A}^{i} \rangle$, we can rewrite Eq.~\eqref{reduced likelihood} as follows,
\begin{equation}
    \mathcal{F} = \frac{1}{2}[S^iM_{ij}S^j + 2S^iM_{ij}N^j + N^iM_{ij}N^j] \,.
\end{equation}
The second term on the right-hand side is the first-order noise term, which follows a Gaussian distribution with a zero mean, and it induces fluctuations of the $\mathcal{F}$-statistic around the signal contribution $S^iM_{ij}S^j$. 
On the other hand, the third term is the second-order noise term. It has a non-zero positive value and becomes non-negligible when the noise level is comparable to the signal. 
This second-order noise term also appears in the estimation of a CGW strain amplitude through Eq.~\eqref{expect value of F} as
\begin{equation}
    h_s^2(f_\mathrm{obs}) \propto E(\mathcal{F}) = E(S^iM_{ij}S^j) + E(N^iM_{ij}N^j).
\end{equation}
This term again has a non-zero positive value and gives a non-negligible contribution when noise levels are high, leading to a non-zero bias.  

The results of \textit{case 1} demonstrate that the performance of the standard $\mathcal{F}$-statistic is degraded when a GWB is present in the data.
This suggestion highlights the need for improved modelling of the $\mathcal{F}$-statistic to account for the presence of GWB.

\subsubsection{Case 2. Performance of the new modelling}
Next, we compare the performance of the standard and GWB-informed modelling of the $\mathcal{F}$-statistic in estimating parameters in the presence of a GWB. The distinction in noise modelling lies in the noise matrix $\Sigma_\mathrm{n}$; the standard modelling is represented as $\Sigma_\mathrm{n, pre} = \Sigma_\mathrm{WN}$, while GWB-informed modelling is expressed as $\Sigma_\mathrm{n, new} = \Sigma_\mathrm{WN} + \Sigma_\mathrm{GWB}$. We apply the $\mathcal{F}$-statistic using these different noise models to mock datasets that include white noise, a single CGW, and a GWB with parameters $A_g = 1.0 \times 10^{-15}$ and $\alpha_g = -2/3$.

\begin{figure*}
\centering
\begin{minipage}[b]{\columnwidth}
    \centering
    \includegraphics[width=0.99\columnwidth]{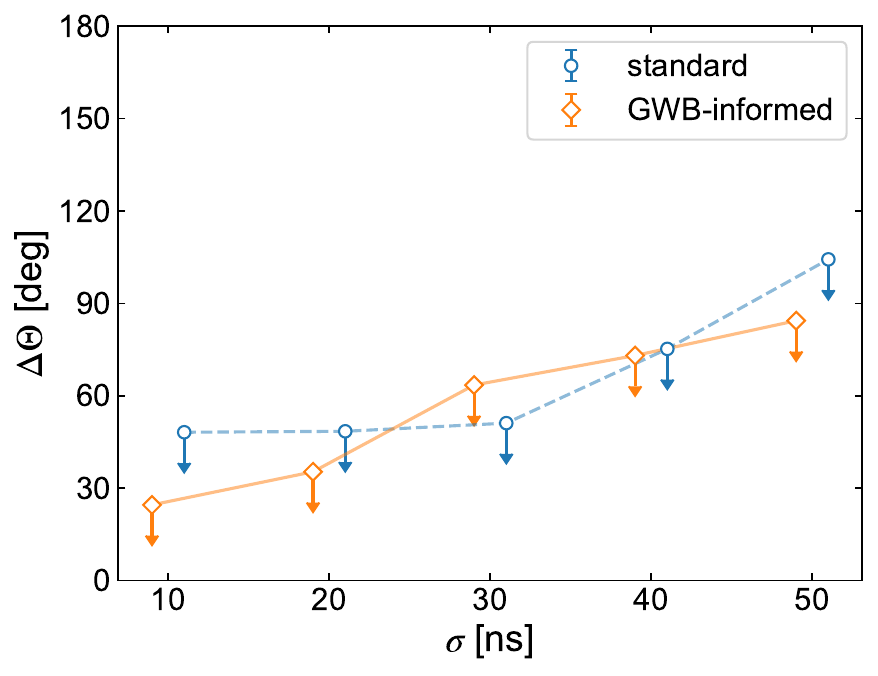}
\end{minipage}
\begin{minipage}[b]{\columnwidth}
    \centering
    \includegraphics[width=0.99\columnwidth]{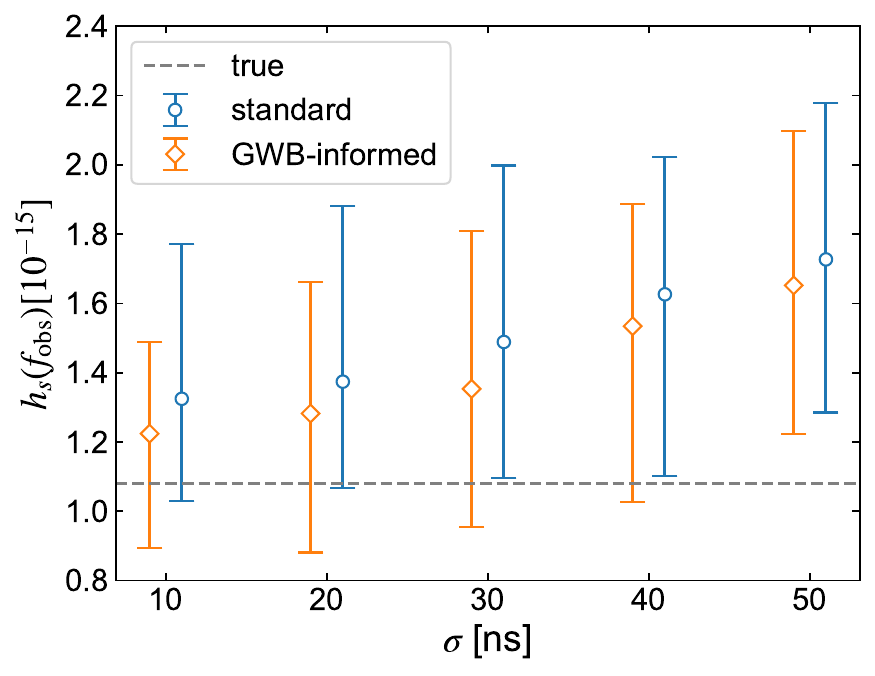}
\end{minipage}
\caption{Results of \textit{case 2}. Predicted parameter values obtained from 100 realizations by applying the standard $\mathcal{F}$-statistic (blue circle with error bars or downward arrows) and GWB-informed $\mathcal{F}$-statistic (red diamond median with error bars or downward arrows). The mock datasets include white noise, a single CGW, and GWB.
The left panel shows the deviation of the source position from the true position of the injected CGW source, evaluated at different noise levels. 
Each point with a downward arrow means that the deviation of the angle is within the indicated value for $68\%$ of the realizations.
The right panel shows the estimated GW strain amplitude of the injected CGW source for different noise levels. The gray line represents the true GW strain of the injected signal. Each point and error bar shows the median and the $16\%$-$84\%$ region of each distribution, respectively.
}
\label{fig:case2}
\end{figure*}

\begin{figure*}
\centering
\begin{minipage}[b]{\columnwidth}
    \centering
    \includegraphics[width=0.99\columnwidth]{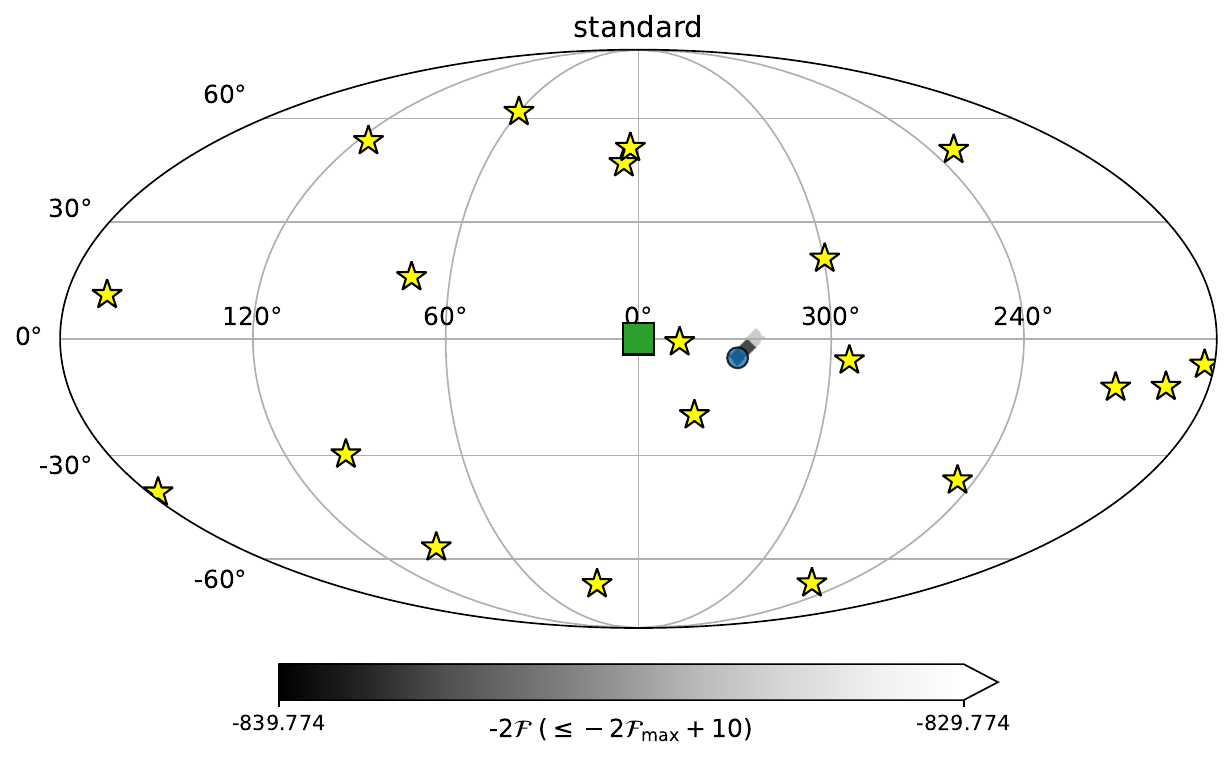}
\end{minipage}
\begin{minipage}[b]{\columnwidth}
    \centering
    \includegraphics[width=0.99\columnwidth]{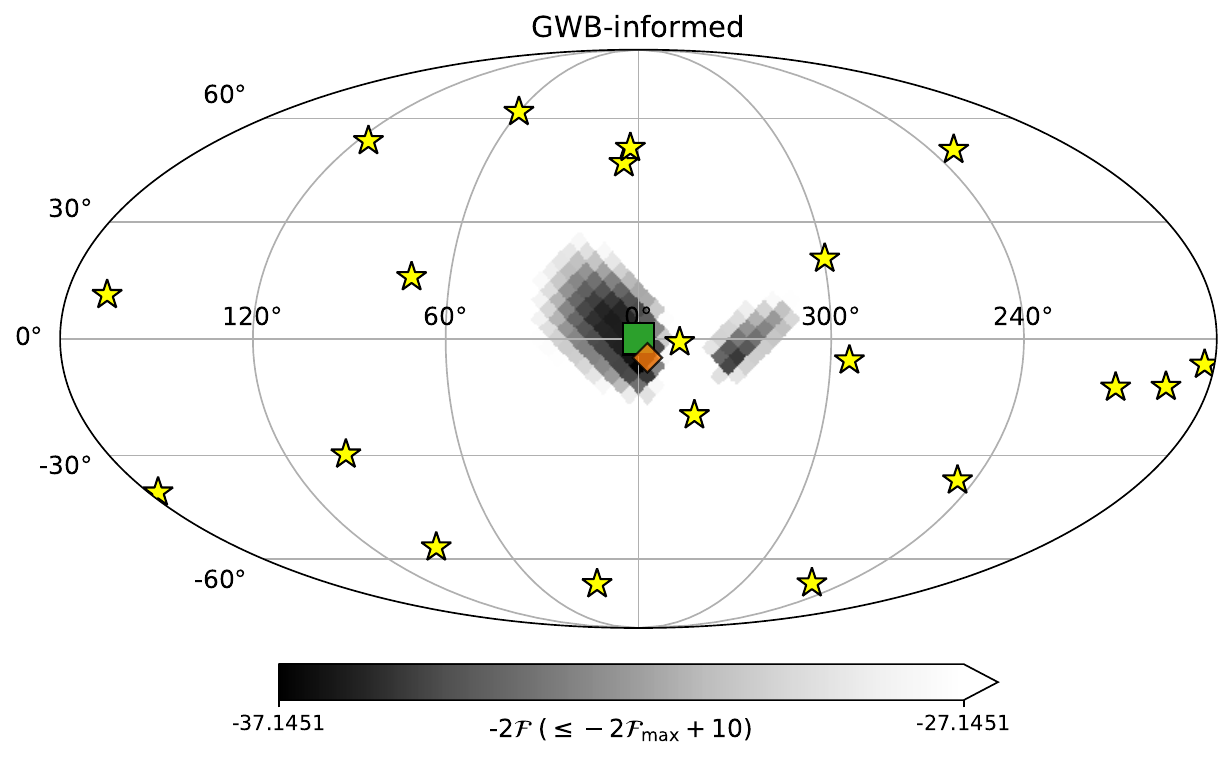}
\end{minipage}
\caption{An example of the skymap of the reduced log-likelihood ratio $\mathcal{F}$ is shown for the standard (left panel) and GWB-informed (right panel) $\mathcal{F}$-statistics in \textit{case 2}, which includes both the GWB and white noise with $\sigma=10$ ns.
The stars and the green square represent the sky positions of pulsars and the injected CGW source, respectively. The blue circle (in the right panel) and the orange diamond (in the left panel) show the positions with the maximum value of $\mathcal{F}=\mathcal{F}_\mathrm{max}$
These heatmaps show the magnitude of $-2\mathcal{F}$ at each position within the region $-2\mathcal{F}\leq-2\mathcal{F}_\mathrm{max}+10$, which corresponds to $\Delta\chi^2 \leq 10$}
\label{fig:case2_skymap}
\end{figure*}

The distributions of the estimated
parameter values are shown in Figure~\ref{fig:case2}. 
The left panel compares the predictions of $\Delta\Theta$ between the standard $\mathcal{F}$-statistic and the GWB-informed one.
As the noise increases, the contribution of the GWB in the noise matrix becomes negligible, and the differences in the accuracy of the estimation between the standard and GWB-informed models become small.
On the other hand, in the case with a low level of white noise where the GWB dominates the noise component, our results show that the GWB-informed $\mathcal{F}$-statistic can more accurately determine the sky position of the CGW compared to the standard model.

The right panel shows the distribution of the predicted GW strain amplitude for both the standard and GWB-informed $\mathcal{F}$-statistic. Our results indicate that the GWB-informed $\mathcal{F}$-statistic can improve the estimation of the GW strain, alleviating the uncertainty and the tendency of overestimation. This demonstrates that the GWB-informed $\mathcal{F}$-statistic can effectively separate the GW component into an individual deterministic CGW and a stochastic GWB.

Furthermore, this modelling also helps to reduce the bias in estimating the CGW source position.
Figure~\ref{fig:case2_skymap} shows an example of the log-likelihood skymap in \textit{case 2} where $\sigma=10$ ns. The standard $\mathcal{F}$-statistic, displayed on the left, yields a narrow uncertainty around the estimated sky position but indicates an incorrect location. In contrast, the GWB-informed $\mathcal{F}$-statistic provides the correct position with a more reasonable level of uncertainty, as demonstrated on the right.

Our results in \textit{case 2} suggest that, even with the presence of the dominant GWB signal, the $\mathcal{F}$-statistic retains its ability to identify the CGW component when employing the GWB-informed noise modelling.

In Appendix~\ref{subsec:appendix.A1}~and~\ref{subsec:appendix.A2}, we further assess the performance of the standard and GWB-informed $\mathcal{F}$-statistic by varying the amplitudes of the GWB and CGW, aiming to numerically identify the conditions for detecting the CGW signal. Furthermore, in Appendix~\ref{subsec:appendix.A3}, we evaluate the performance of the GWB-informed $\mathcal{F}$-statistic when we use a different noise covariance matrix of GWB from the one we used to realise GWB in our simulation.

\subsection{More realistic case}\label{subsec:realistic cases}
In this subsection, we apply the standard and GWB-informed $\mathcal{F}$-statistic to more realistic setups by generating mock data based on a phenomenological SMBH population model.
Here we adopt two scenarios for the PTA setup by following \cite{2012PhRvD..85d4034B}: \textit{Hard search} ($N_\mathrm{p}=30$ pulsars with equal white noise levels of $\sigma=100$ $\mathrm{ns}$) and \textit{Easy search} ($N_\mathrm{p}=50$ pulsars with equal white noise levels of $\sigma=30$ $\mathrm{ns}$). The former setup is similar to current PTA observations, whereas the latter considers the PTA observations in the SKA era.
We fix the distribution of pulsars on the sky map while changing the parameters of CGWs and the sampling of white noise and GWB in each realization.

We generate CGWs and GWB using the Agnostic Model, a phenomenological model introduced in \cite{2016MNRAS.455L..72M}. The Agnostic Model describes the comoving merger rate of SMBH binaries per unit redshift $z$ and unit chirp mass $M_c$ as the analytical formula characterised by five hyperparameters, which is given by
\begin{equation}
\begin{split}
\label{AgnosticModel}
    \frac{\dd^3N}{\dd V_\mathrm{c}\dd z\dd\log_{10}M_c} &= \dot{n}_0\Big(\frac{M_c}{10^7M_\odot}\Big)^{-\alpha_\star}\ee^{-M_c/M_\star}\\
    &\times(1+z)^{\beta_\star}\ee^{-z/z_\star}\frac{\dd t_\mathrm{r}}{\dd z}~,
\end{split}
\end{equation}
where $V_\mathrm{c}$ denotes a comoving volume, $t_\mathrm{r}$ is the time measured in the rest frame of the GW source at redshift $z$, $\dot{n}_0$ represents the normalised comoving merger rate, and 
\begin{equation}
    \frac{\dd t_\mathrm{r}}{\dd z} = \frac{1}{(1+z)H_0\sqrt{\Omega_\mathrm{m}(1+z)^3+\Omega_{\Lambda}}}.
\end{equation}
The parameters $\beta_\star$ and $z_\star$ control the redshift distribution of the merger rate, while $\alpha_\star$ and $M_\star$ determine the chirp mass distribution.
The merger rate per unit redshift $z$, logarithmic chirp mass $M_c$, and logarithmic frequency is calculated by
\begin{equation}
\label{AgnosticModel_1}
    \frac{\mathrm{d}^3N}{\mathrm{d}z\mathrm{d}\log_{10}M_c\mathrm{d}\ln f} = \frac{\dd^3N}{\dd V_\mathrm{c}\dd z\dd\log_{10}M_c} \frac{\dd V_\mathrm{c}}{\dd t_\mathrm{r}}\frac{\dd t_\mathrm{r}}{\dd \ln f_r} \,.
\end{equation}
Note that $\dd\ln f_\mathrm{r} = \dd\ln [(1+z)f] = \dd\ln f $. The change in the comoving volume per rest-frame time $t_\mathrm{r}$ is 
\begin{equation}
    \frac{\dd V_\mathrm{c}}{\dd t_\mathrm{r}} = 4\pi c \frac{d_L^2}{(1+z)^2} \,,
\end{equation}
and assuming all SMBH binaries are inspiralling and have circular orbits, we have
\begin{equation}
\frac{\dd t_\mathrm{r}}{\dd \ln f_\mathrm{r}} = \frac{5}{96}\Big(\frac{GM_c}{c^3}\Big)^{-5/3}(1+z)^{-8/3}(\pi f)^{-8/3} \,.
\end{equation}

\begin{figure}
\centering
\includegraphics[width=\columnwidth]{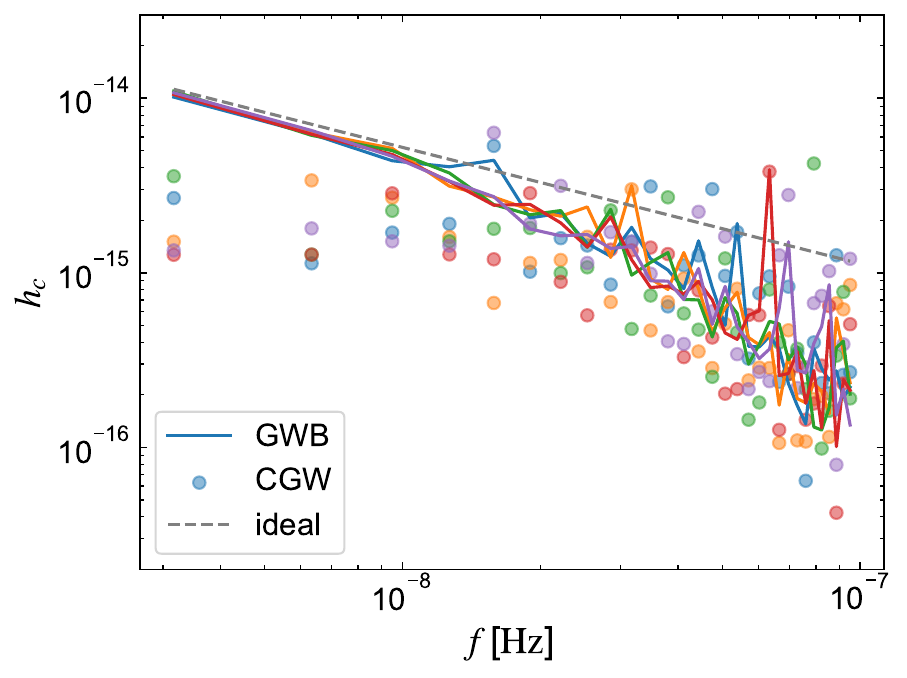}
\caption{The GW spectrum is numerically computed using the SMBH population from the Agnostic Model. Five different colours represent five different realizations. The circles indicate the loudest GW in each frequency bin, and the lines represent the assembly of unresolved GW. The Gray dashed line shows the power-law spectrum with $\alpha_g=-2/3$, which is normalised to match the estimate from the latest NANOGrav's observation \citep{2023ApJ...951L...8A}.}
\label{fig:AG_spectrum}
\end{figure}

Using the merger rate given in Eq.~\eqref{AgnosticModel_1} and summing all squared GW strains of inspiralling SMBHs given in Eq.~\eqref{single GW strain}, the GW spectrum is given by
\begin{equation}
\label{GWBspec_from_SMBH}
\begin{split}
    h_{c}^2(f) &= \int_{0}^{\infty}\dd z \int\dd \log_{10} M_c \frac{\dd^3 N}{\dd z \dd \log_{10} M_c \dd \ln f} h_s^2(f)\\
    &= \frac{4G^{5/3}}{3c^2\pi^{1/3}}f^{-4/3}\int_{0}^{\infty}\frac{\dd z}{(1+z)^{1/3}}\\
    &\times \int \dd \log_{10} M_c M_c^{5/3} \frac{\dd^3 N}{\dd V_\mathrm{c} \dd \log_{10} M_c \dd z} \,.\\
\end{split}
\end{equation}
The GWB spectrum obtained in this way follows a power-law spectrum with $\alpha_g = -2/3$. We choose the values of the five hyperparameters as ($\dot{n}_0$, $\alpha_\star$, $M_\star$, $\beta_\star$, $z_\star$) = ($10^{-4.45}\mathrm{Mpc}^{-1}\mathrm{Gyr}^{-1}$, $0$, $1.0\times10^9M_\odot$, $2.5$, $2.5$) to reproduce the amplitude of GWB consistent with the estimate of the latest observation of NANOGrav, namely, $A_g = 2.4\times10^{-15}$ for $\alpha_g = -2/3$.

\begin{figure*}
\centering
\begin{minipage}[b]{\columnwidth}
    \centering
    \includegraphics[width=0.99\columnwidth]{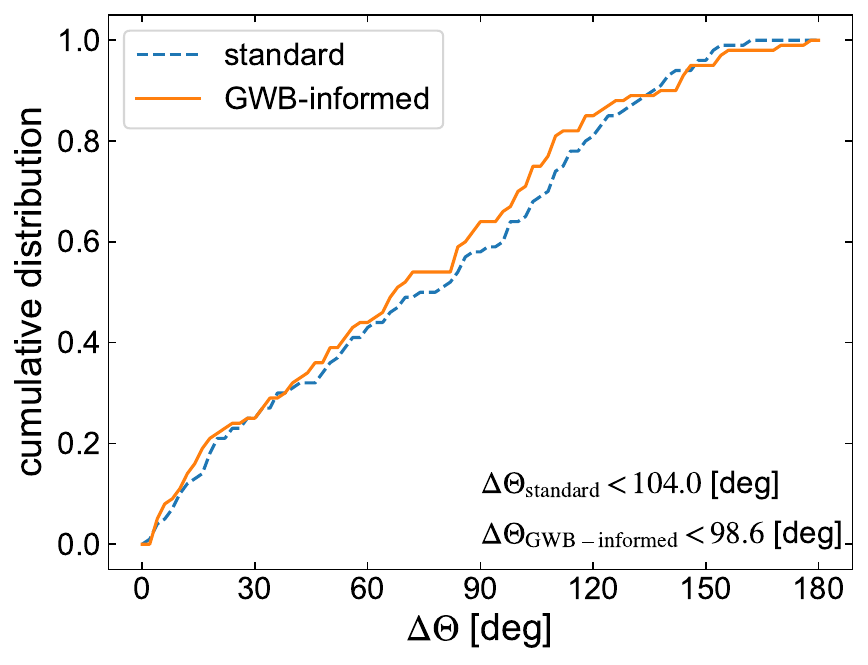}
\end{minipage}
\begin{minipage}[b]{\columnwidth}
    \centering
    \includegraphics[width=0.99\columnwidth]{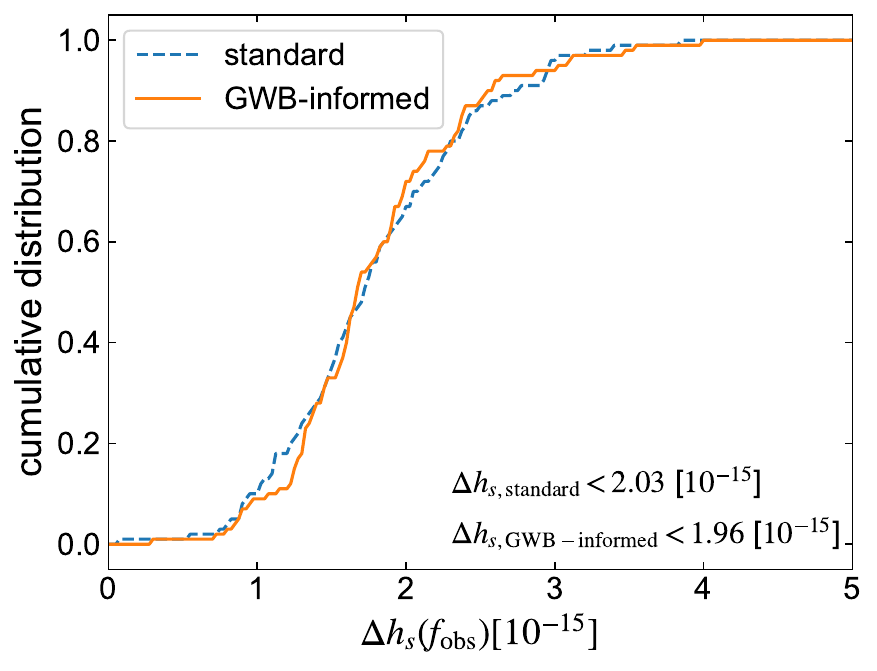}
\end{minipage}
\caption{Results of \textit{Hard search} ($N_\mathrm{p}=30$, $\sigma=100$ $\mathrm{nHz}$). The cumulative error distribution of the predictions was statistically evaluated over 100 realizations. The left panel shows the direction error of the injected CGW source, and the right panel shows the error of the GW strain. The blue dashed line and the orange solid line represent the results estimated by the standard and GWB-informed $\mathcal{F}$-statistic, respectively.
The $68\%$ values obtained for the two modellings are indicated at the bottom right of each figure.
}
\label{fig:AG_hard}
\end{figure*}

\begin{figure*}
\centering
\begin{minipage}[b]{\columnwidth}
    \centering
    \includegraphics[width=0.99\columnwidth]{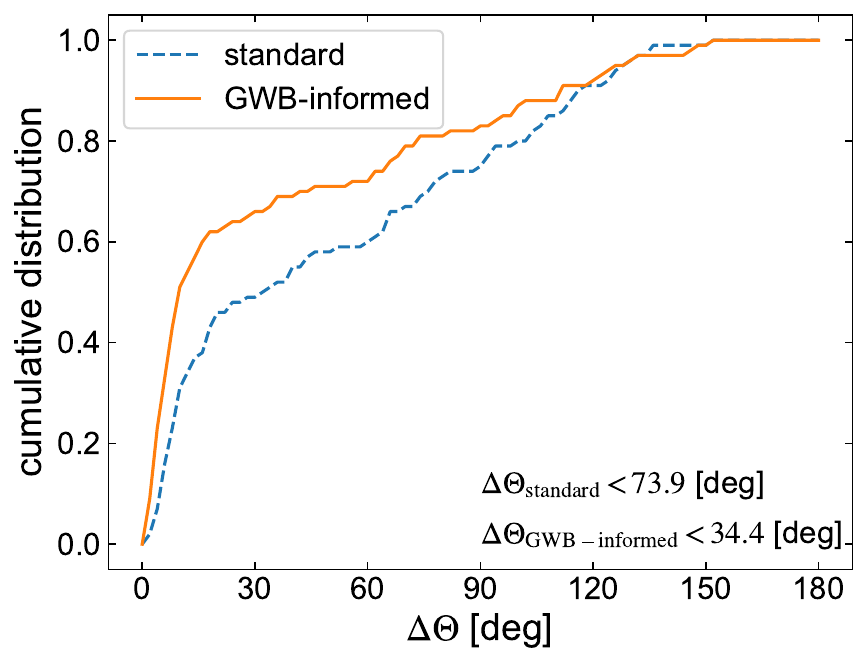}
\end{minipage}
\begin{minipage}[b]{\columnwidth}
    \centering
    \includegraphics[width=0.99\columnwidth]{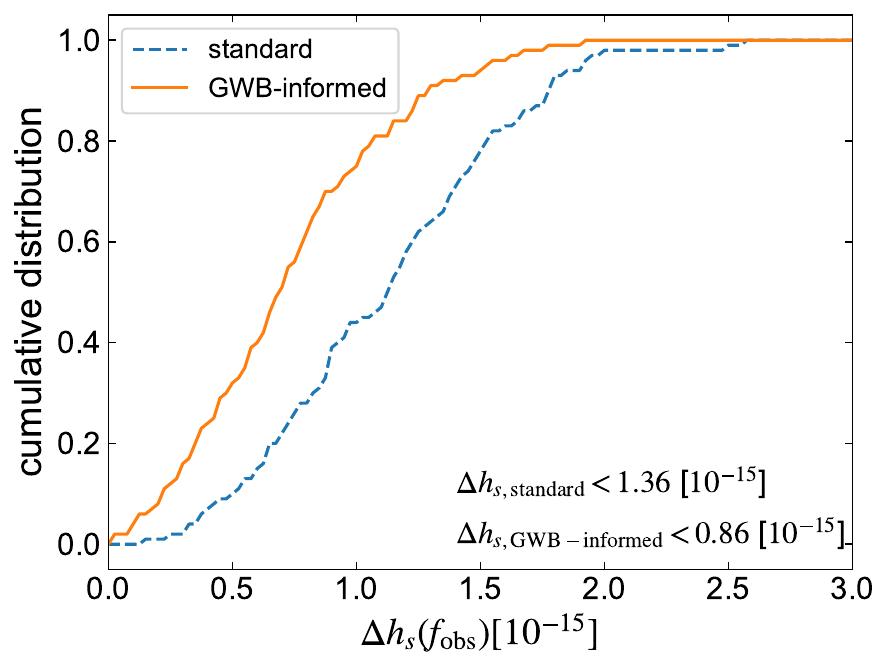}
\end{minipage}
\caption{Results of \textit{Easy search} ($N_\mathrm{p}=50$, $\sigma=30$ $\mathrm{nHz}$). Shown is the same as Figure \ref{fig:AG_hard}.}
\label{fig:AG_easy}
\end{figure*}

The merger rate given by Eq.~\eqref{AgnosticModel_1} is the expected value, whereas, in reality, GWs originate from a discrete number of SMBH binaries. To reproduce GWs observed in real observations, we numerically generate the number of SMBH mergers using Poisson sampling~\citep{2003ApJ...583..616J,2024ApJ...965..164G},
\begin{equation}
\label{poisson sampling}
    N(z,\mathcal{M}_c,f) = \mathcal{P}\Big(\frac{\mathrm{d}^3N}{\mathrm{d}z\mathrm{d}\log_{10}M_c\mathrm{d}\ln f}\Delta z \Delta \log_{10}M_c\Delta\ln f\Big) \,,
\end{equation}
and by summing up the squared GW strains according to the number of SMBH mergers Eq.~\eqref{poisson sampling}, we obtain
\begin{equation}
    h_c^2(f) = \sum_{z, M_c}N(z, M_c, f)h_s^2(f)\Big(\frac{f}{\Delta f}\Big).
\end{equation}

Figure~\ref{fig:AG_spectrum} plots the predicted GW spectrum from SMBHs based on the Agnostic Model, where each event is randomly generated by the Poisson sampling. To generate mock timing residuals, we first select the brightest GW source in each frequency bin (represented as circles in the figure) and inject it as a CGW. The remaining sources are approximated as a GWB with a broken power-law spectrum \citep{2008MNRAS.390..192S} with the function form of
\begin{equation}
    h_{\mathrm{GWB}, c}(f) = A_g\Big(\frac{f}{f_\mathrm{ref}}\Big)^{-2/3}\Big(1+\frac{f}{f_0}\Big)^{\gamma_0}.
\end{equation}
The parameters ($f_0$, $\gamma_0$) are determined by fitting the averaged unresolved GW component (shown as coloured solid lines) over 100 realizations, and we find ($f_{0}$, $\gamma_0$) = ($35$ $\mathrm{nHz}$, $-1.0$).

Figures~\ref{fig:AG_hard} and~\ref{fig:AG_easy} show the cumulative error distribution of the predictions in \textit{Hard search} and \textit{Easy search}, respectively, obtained by the standard and GWB-informed $\mathcal{F}$-statistic. 
In \textit{Hard search}, corresponding to the current PTA setup, the GWB contribution is not dominant compared to the white noise.
Therefore, these two modellings of $\mathcal{F}$-statistic produce almost the same results. 
On the other hand, in \textit{Easy search}, corresponding to the future PTA setup in the SKA era, the contribution of GWB exceeds the white noise level. In this case, the GWB-informed $\mathcal{F}$-statistic has smaller errors in both the sky position and the GW strain in this likely scenario.
Thus, our results from these mock data simulations suggest that the previous modelling is currently valid but will no longer be applicable in future PTA observations. By introducing the new modelling, the $\mathcal{F}$-statistic will be capable of resolving an individual CGW in the anticipated scenarios.

\section{summary \& discussion}\label{sec:summary&discussion}
In future PTA measurements, individual CGWs and the GWB are expected to be observed simultaneously. To develop algorithms and pipelines suited for this anticipated scenario in the SKA era, we have revisited the application of the $\mathcal{F}$-statistic in PTA observations, a detection method designed for single-source signals. The $\mathcal{F}$-statistic provides an efficient maximum likelihood estimate by analytically reducing the number of unknown parameters of a GW template. Additionally, it can be easily extended to handle multiple CGWs.  As a first step, we evaluated its performance using simple mock datasets that include both a single CGW and a GWB.

Our results demonstrate that the presence of unresolved GWs, when dominating the white noise, indeed introduces a bias in parameter estimation. Specifically, in the presence of a GWB, the analysis tends to infer incorrect source positions with narrow uncertainties and also overestimates the strain amplitude.
Our new modelling of the $\mathcal{F}$-statistic, which treats unresolved GWs as a noise component, can mitigate biases in parameter estimation and accurately recover the parameters of the injected CGW. Additionally, we applied both the standard and GWB-informed $\mathcal{F}$-statistics to more realistic scenarios based on an ideal SMBH population model. The results showed no significant difference in parameter estimation between the two methods under current PTA observations. However, we demonstrated that in the expected PTA configuration of the SKA era, it is essential to use the GWB-informed $\mathcal{F}$-statistic, as it provides more accurate parameter predictions than the standard approach.

Let us discuss the potential for developing an algorithm based on the $\mathcal{F}$-statistic. When we calculate the $\mathcal{F}$-statistic, we assume prior knowledge of the noise components, including GWB. This assumption is essential because we aim to use the statistical quantity $\mathcal{F}$ to accurately identify individual CGWs in the presence of GWB.
However, in the real PTA observations, we need to estimate the parameters of both CGWs and GWB simultaneously (see e.g., \citealt{2023ApJ...959....9B} and \citealt{2024arXiv240721105F}). 
Therefore, we should optimise not just the log-likelihood ratio $\ln\Lambda$, but the full log-likelihood, which is given as follows:
\begin{equation}
    \ln\mathcal{L} = -\frac{1}{2}\ln\mathrm{det}(2\pi\Sigma_\mathrm{n}) - \frac{1}{2}\langle\vb{x}|\vb{x}\rangle + \ln\Lambda \,.
\end{equation}
As discussed in Section~\ref{subsec:simple cases}, we have confirmed that the reduced log-likelihood $\mathcal{F}$ works well even in the presence of GWB. Therefore, we propose adopting the alternative log-likelihood based on the $\mathcal{F}$-statistic instead of using the full log-likelihood, 
\begin{equation}
    \ln\mathcal{L}_{\mathcal{F}} = -\frac{1}{2}\ln\mathrm{det}(2\pi\Sigma_\mathrm{n}) - \frac{1}{2}\langle\vb{x}|\vb{x}\rangle + \mathcal{F},
\end{equation}
which helps to reduce the number of parameters. We aim to develop an algorithm to efficiently explore the likelihood, including GWB parameters, in future work.

Another challenge is that, in future PTA measurements with higher sensitivities, we need to consider the possibility of multiple CGWs existing within one frequency bin. In this case, we generally have to deal with multimodal parameter estimation. For this, we need to establish the algorithm based on the combination of $\mathcal{F}$-statistic and the proper optimization algorithm, for example, generic algorithm \citep{2013PhRvD..87f4036P}, MCMC, and MultiNest \citep{2009CQGra..26u5003F}. 

The SKA project is expected to enable highly sophisticated GW observations through the combined timing residuals of $10^{3}$–$10^{4}$ pulsars.  Under this scenario, assuming $T_\mathrm{obs} = 10$ $\mathrm{yr}$ and $\Delta t = 2$ $\mathrm{week}$,
the size of the noise matrix increases from $10^5\times10^5$ to $10^6\times10^6$. When maximizing the likelihood, calculating the inverse of this large noise matrix at each iteration presents a significant computational challenge. To address this issue, \cite{2013ApJ...769...63E} introduced a method to reduce computational costs by expanding the noise matrix in terms of the GWB amplitude, under the assumption of a strong white noise limit.

Despite the challenges outlined above, the $\mathcal{F}$-statistic still holds significant potential as a foundation for developing pipelines to explore new physics in future PTA observations.

\section*{acknowledgments}
KF is supported by Japan Society for the Promotion of Science (JSPS) KAKENHI Grant Number JP25KJ1388.
KI is supported in part by the JSPS grant numbers 21H04467, 24K00625, JST FOREST Program JPMJFR20352935, and the JSPS Core-to-Core Program (grant numbers: JPJSCCA20200002, JPJSCCA20200003). SK is partly supported by the I+D grant PID2023-149018NB-C42, the Consolidaci\'on Investigadora 2022 grant CNS2022-135211, and the Grant IFT Centro de Excelencia Severo Ochoa No CEX2020-001007-S, funded by MCIN/AEI/10.13039/501100011033, the Leonardo Grant for Scientific Research and Cultural Creation 2024 from the BBVA Foundation, and Japan Society for JSPS KAKENHI Grant no.  JP23H00110 and JP24K00624.
%\newpage

\section*{Data Availability}
The data underlying this paper will be shared on reasonable request to the corresponding author.

\bibliography{ref}{}
\bibliographystyle{mnras}

\appendix

\section{Other simple cases}\label{sec:appendix.A}
In addition to the simple cases investigated in Section~\ref{subsec:simple cases}, in this appendix, we examine three different cases to evaluate the performance of $\mathcal{F}$-statistic. The basic setup of the mock data simulations here is the same as in Section~\ref{sec:results}.
We consider different amplitudes of GWB or the strain of the injected CGW from \textit{Case 1} and \textit{Case 2}.  We define $\lambda_\mathrm{fore}\equiv h_{\mathrm{CGW},c}^2/(h_{\mathrm{CGW},c}^2+h_{\mathrm{GWB},c}^2)$ to quantify the amplitudes of the injected CGW and GWB. This quantity was introduced by \cite{2018MNRAS.477..964K} and represents the fraction of the GW energy from the injected CGW against the total GW energy at $f=f_\mathrm{obs}$.
In the previous work, $\lambda_\mathrm{fore}=0.5$ was chosen to be the fiducial threshold for the detection of a CGW. 
Here we evaluate the performance of $\mathcal{F}$-statistic for various values $\lambda_\mathrm{fore}$ that helps determine the detection threshold for a single CGW when adapting the $\mathcal{F}$-statistic.

\subsection{Case 3: Varying the CGW amplitude while keeping the GWB amplitude constant}\label{subsec:appendix.A1}
\begin{figure*}
\centering
\begin{minipage}[b]{\columnwidth}
    \centering
    \includegraphics[width=0.99\columnwidth]{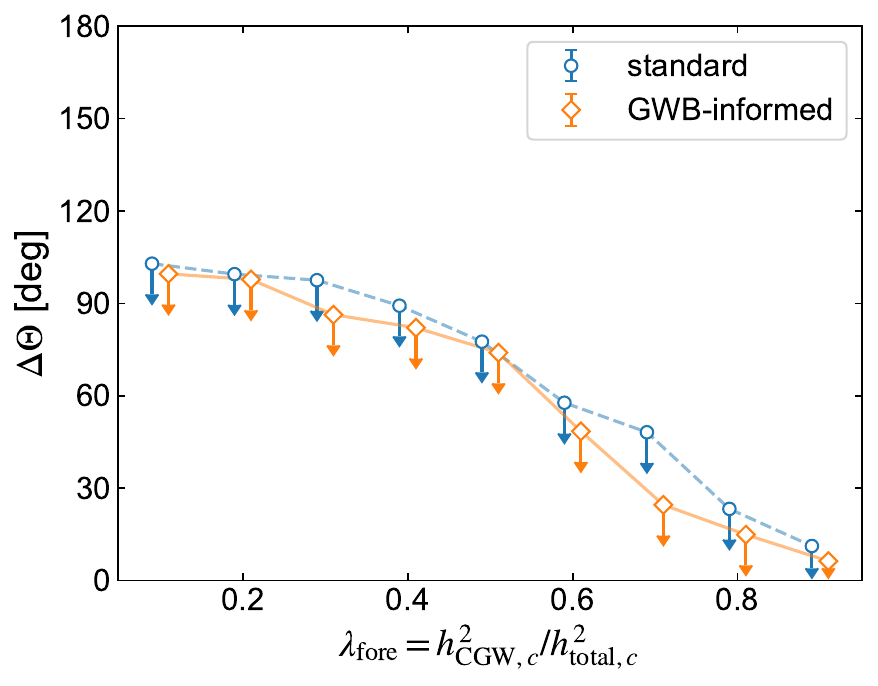}
\end{minipage}
\begin{minipage}[b]{\columnwidth}
    \centering
    \includegraphics[width=0.99\columnwidth]{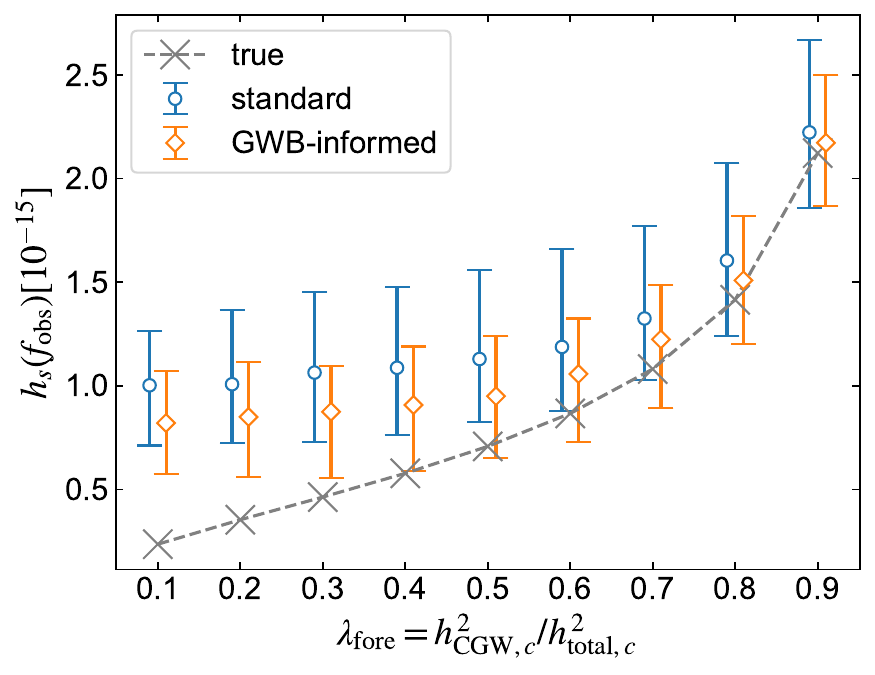}
\end{minipage}
\caption{Results of \textit{case 3}. Predicted values of source direction (left panel) and strain amplitude (right panel) over 100 realizations are shown. This analysis utilised both the standard $\mathcal{F}$-statistic (represented by blue circles with error bars or downward arrows) and the GWB-informed $\mathcal{F}$-statistic (represented by red diamonds with error bars or downward arrows) on mock data sets. These datasets included white noise with a noise level $\sigma=10$ ns, along with a CGW and a GWB characterised by parameters $(A_g, \alpha_g) = (1.0 \times 10^{-15}, -2/3)$. 
Each point with a downward arrow means that the deviation of the angle is within the indicated value for $68\%$ of the realizations.
The right panel shows the estimated GW strain amplitude of the injected CGW source for different $\lambda_\mathrm{fore}$. The gray line represents the true GW strain of the injected signal. Each point and error bar shows the median and the $16\%$-$84\%$ region of each distribution, respectively.}
\label{fig:case3}
\end{figure*}

In \textit{Case 3}, we set the white noise level to $\sigma=10$ ns and fix the parameters of GWB to ($A_g$,$\alpha_g$)=($1.0\times10^{-15}$, $-2/3$), and vary the strain of the injected CGW. Here, we consider $\lambda_\mathrm{fore} = 0.1,0.2,...,0.9$ and apply both standard and GWB-informed $\mathcal{F}$-statistics on the mock datasets consisting of these signals.
The results in \textit{Case 3} are shown in Figure~\ref{fig:case3}.
As shown in the left panel, the GWB-informed $\mathcal{F}$-statistic can estimate the source direction more accurately than the standard $\mathcal{F}$-statistic for $\lambda_\mathrm{fore} \geq 0.7$. Specifically, for $\lambda_\mathrm{fore} \geq 0.7$, the GWB-informed $\mathcal{F}$-statistic can pinpoint the direction within $\Delta\Theta \leq 30^\circ$.
The right panel shows that the GWB-informed $\mathcal{F}$-statistic also estimates the CGW strain more precisely than the standard approach across all values of $\lambda_\mathrm{fore}$. This overall trend is due to the GWB-informed $\mathcal{F}$-statistic’s ability to mitigate the influence of the GWB by modelling it as a noise component. For $\lambda_\mathrm{fore} \geq 0.5$, the GWB-informed $\mathcal{F}$-statistic can accurately estimate the CGW strain within the $16\%$-$84\%$ range.

\subsection{Case 4: Varying the GWB amplitude while keeping the CGW amplitude constant}\label{subsec:appendix.A2}
\begin{figure*}
\centering
\begin{minipage}[b]{\columnwidth}
    \centering
    \includegraphics[width=0.99\columnwidth]{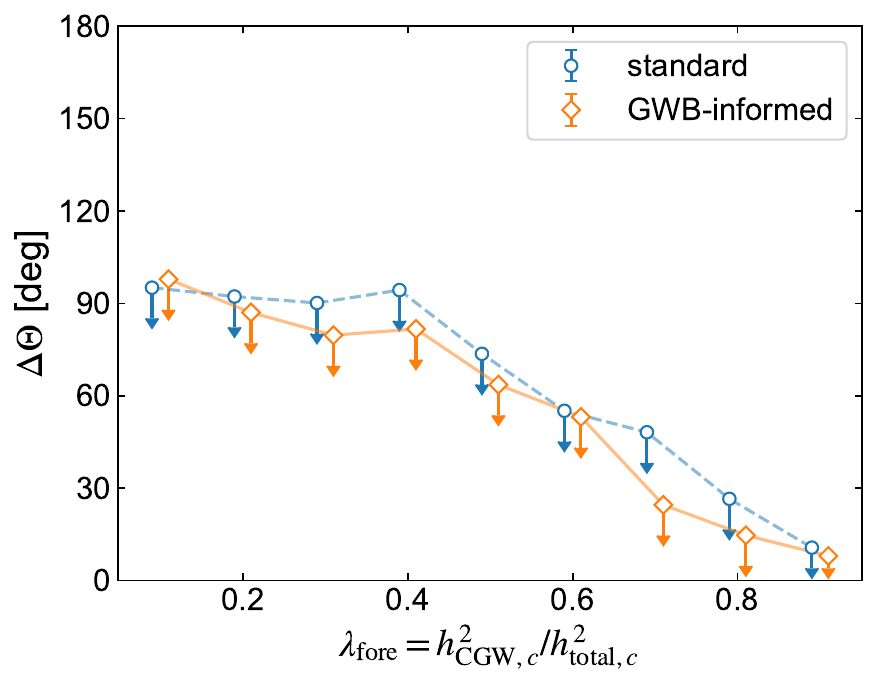}
\end{minipage}
\begin{minipage}[b]{\columnwidth}
    \centering
    \includegraphics[width=0.99\columnwidth]{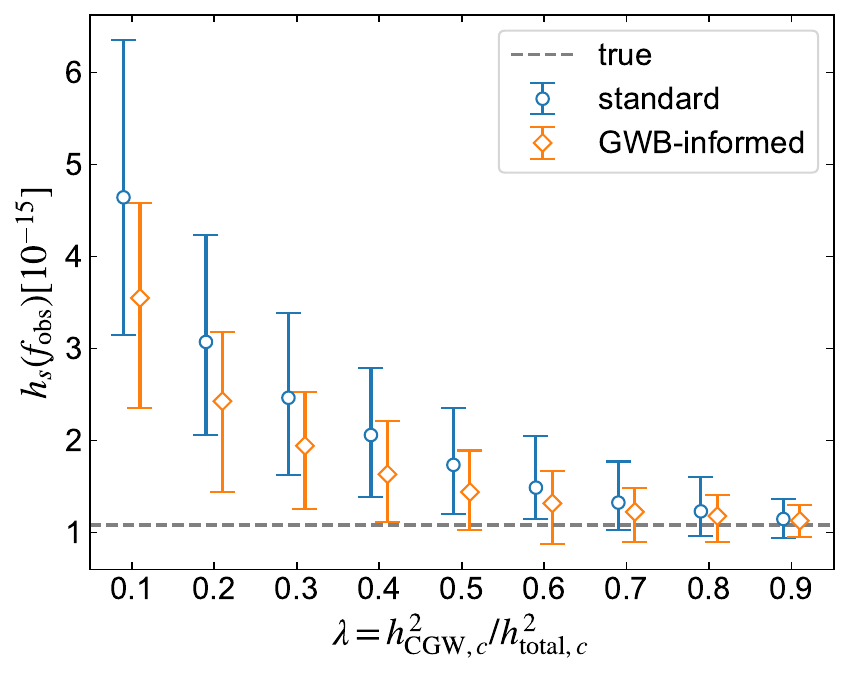}
\end{minipage}
\caption{Results of \textit{case 4}. 
Shown is the same as Figure \ref{fig:case3}, but 
in this case, the strain of the injected CGW is fixed to $h_\mathrm{s}(f_\mathrm{obs})=1.1\times 10^{-15}$, while varying the amplitude of GWB.
The left panel shows the direction errors of the injected CGW source for different $\lambda_\mathrm{fore}$. The right panel shows the estimated GW strain of the injected CGW source for different $\lambda_\mathrm{fore}$. The gray line represents the true GW strain of the injected signal. }
\label{fig:case4}
\end{figure*}

In \textit{Case 4}, we fix the CGW strain to $h_s(f_\mathrm{obs})=1.1\times 10^{-15}$, but vary the amplitude of GWB with $\lambda_\mathrm{fore}=0.1,0.2,...,0.9$. The results in \textit{Case 4} are shown in Figure~\ref{fig:case4}. The conclusion drawn from these results is nearly identical to that from \textit{Case 3}. 
Based on our numerical simulations in \textit{case 3} and \textit{case 4}, we may set a conservative detection threshold of $\lambda_\mathrm{fore} = 0.7$ for the GWB-informed $\mathcal{F}$-statistic.

\subsection{\textit{Case 5}: When the GWB amplitude is unknown}\label{subsec:appendix.A3}
\begin{figure*}
\centering
\begin{minipage}[b]{\columnwidth}
    \centering
    \includegraphics[width=0.99\columnwidth]{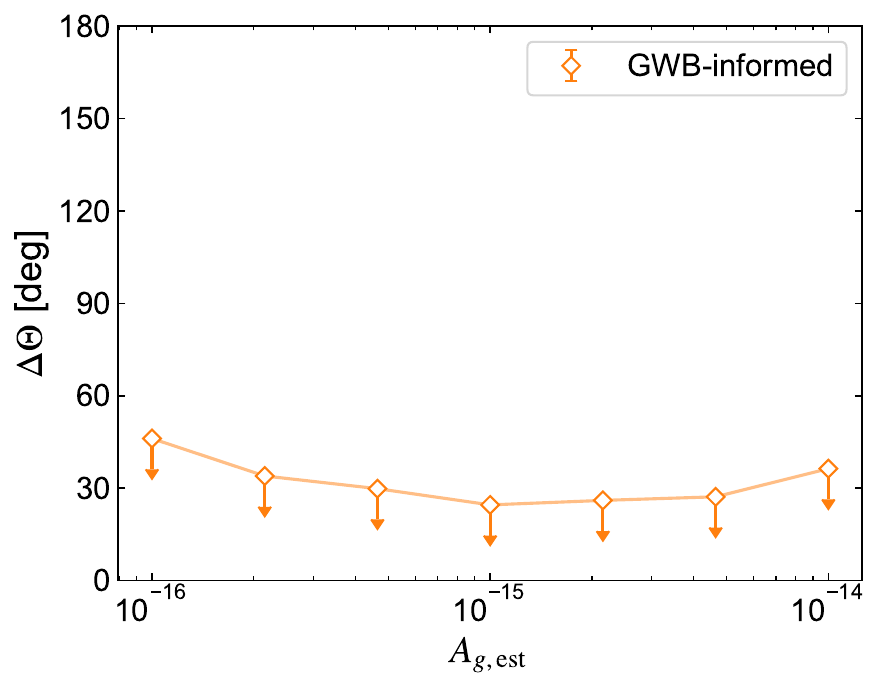}
\end{minipage}
\begin{minipage}[b]{\columnwidth}
    \centering
    \includegraphics[width=0.99\columnwidth]{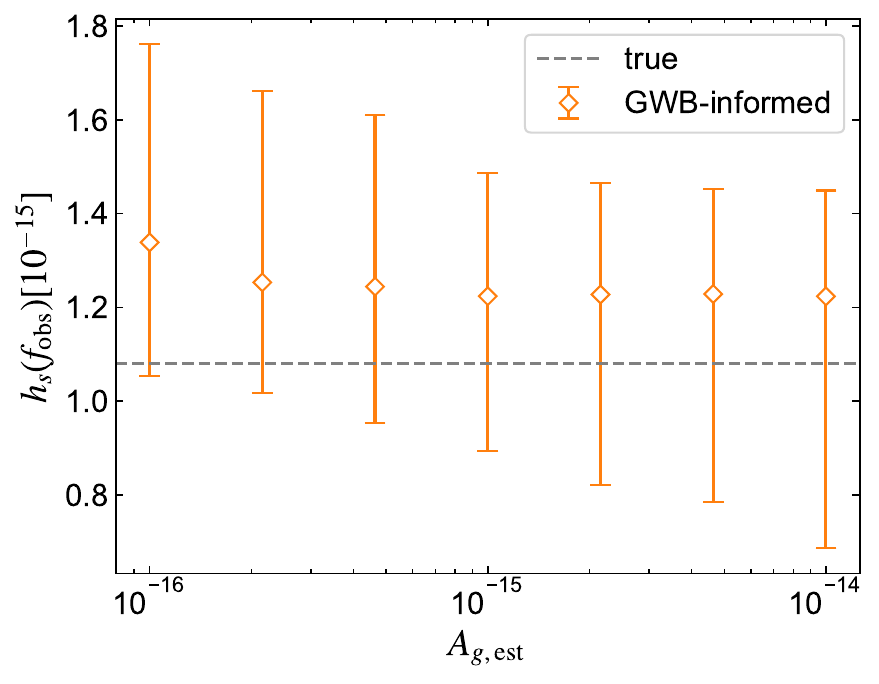}
\end{minipage}
\caption{Results of \textit{case 5}. 
Shown is the same as Figure \ref{fig:case3}, but 
in this case, the estimated amplitude of GWB $A_{g, \mathrm{est}}$ in the noise matrix is varied.
The gray line represents the true GW strain of the injected signal.
}
\label{fig:case5}
\end{figure*}
In our simulations, we assumed prior knowledge of the injected noise components, specifically white noise and the GWB. This assumption enabled us to focus on comparing the estimates of the injected CGW obtained using different modelling approaches within the $\mathcal{F}$-statistic. However, this assumption does not reflect the reality of PTA observations. To address this limitation, we examine a more realistic scenario where the assumed amplitude of the GWB prior to applying the $\mathcal{F}$-statistic differs from the actual injected GWB amplitude. 
In this case, the dataset includes a CGW source with $h_s(f_\mathrm{obs})=1.1 \times 10^{-15}$, white noise with $\sigma=10$ ns, and a GWB characterised by ($A_g$, $\alpha_g$) = ($1.0 \times 10^{-15}$, $-2/3$).
We calculate our noise matrix assuming a GWB amplitude $A_{g,\mathrm{est}}$ that differs from the true value when applying the GWB-informed $\mathcal{F}$-statistic.

Figure~\ref{fig:case5} presents the results for \textit{Case 5}. The left panel shows that the errors in estimating the CGW source direction increase slightly when $A_{g, \mathrm{est}}$ values deviate from the injected value of $A_{g, \mathrm{est}}=1.0 \times 10^{-15}$.
The right panel suggests that the GWB-informed $\mathcal{F}$-statistic estimates a larger CGW strain when a smaller estimated GWB amplitude is used. However, if the estimated amplitude of GWB is taken as $A_{g,\mathrm{est}}\geq A_{g}$, the uncertainty in the CGW strain estimate extends toward smaller values. This behaviour occurs because the contribution of the CGW in the dataset is absorbed by an overestimated GWB component. Consequently, the GWB-informed $\mathcal{F}$-statistic yields different CGW strain predictions depending on the prior estimate of the GWB amplitude.

% Don't change these lines
\bsp	% typesetting comment
\label{lastpage}
\end{document}